\documentclass{aa}  
\usepackage{natbib}
\bibpunct{(}{)}{;}{a}{}{,} 

\usepackage{subfigure}
\usepackage{caption}
\usepackage{xcolor}
\usepackage{array}
\usepackage{graphicx}
\usepackage{tikz}
\usepackage{longtable}
\usepackage{booktabs}
\usepackage{url}
\usepackage{hyperref}
\usepackage{txfonts}
\usepackage{xcolor}
\usepackage[normalem]{ulem}

\begin{document} 

    \title{Constraining reionization morphology and source properties with 21cm galaxy cross-correlation surveys}

   \author{Yannic Pietschke\inst{\ref{inst1}, \ref{inst2}}\fnmsep\thanks{\email{pietschke@thphys.uni-heidelberg.de}} \and
   Anne Hutter\inst{\ref{inst2}} \and
   Caroline Heneka\inst{\ref{inst1}}
          }
    \authorrunning{Pietschke et al.}

   \institute{Institut f\"ur Theoretische Physik, Universit\"at Heidelberg, Philosophenweg 16, 69120 Heidelberg, Germany 
   \label{inst1} 
   \and 
   Institute for Astronomy, University of Vienna, T\"urkenschanzstrasse 17, A-1180 Vienna, Austria \label{inst2}
         }

   \date{Received - -, -; accepted - -, -}
   
\abstract
{Cross-correlations between 21cm observations and galaxy surveys provide a powerful probe of reionization by providing robustness against foreground contamination while linking ionization morphology to galaxies. We quantified the constraining power of 21cm galaxy cross-power spectra for inferring the neutral hydrogen fraction, $x_\mathrm{HI}(z),$ and mean overdensity, $\langle 1+\delta_\mathrm{HI} \rangle(z)$, exploring dependence on the field of view; redshift precision, $\sigma_z$; and minimum halo mass, $M_\mathrm{h,min}$. We employed our simulation-based inference framework \texttt{EoRFlow} for likelihood-free parameter estimation. Mock observations include thermal noise for 100h of SKA-Low with foreground avoidance and realistic galaxy-survey effects. For a fiducial survey ($\mathrm{FOV}=100\,\mathrm{deg}^2$, $\sigma_z=0.001$, $M_\mathrm{h,min}=10^{11}\mathrm{M}_\odot$), cross-power spectra yield unbiased constraints with posterior volumes (PVs) of $\sim$10\% relative to priors. Cross-power measurements reduce the PV by 20--30\% versus 21cm auto-power alone. With foreground avoidance, spectroscopic redshift precision is essential; photometric redshifts render cross-correlations uninformative. Notably, cross-power spectra constrain ionizing source properties, the escape fraction $f_\mathrm{esc,}$ and the star formation efficiency $f_*$, which remain degenerate in auto-power (PV $>$60\%). Tight constraints require either deep surveys detecting faint galaxies 
($M_\mathrm{h,min} \sim 10^{10}\mathrm{M}_\odot$) with moderate foregrounds ($\mathrm{PV}\sim11\%$) or conservative mass limits with optimistic foreground removal ($\mathrm{PV}\sim19\%$). 21cm galaxy cross-correlations enhance morphology constraints beyond auto-power while enabling previously inaccessible source property constraints. Realizing full potential requires precise redshifts and either faint galaxy detection limits or improved 21cm foreground cleaning.
}


   \maketitle

\section{Introduction} \label{sec_introduction}
The epoch of reionization (EoR) marks the transformational period in cosmic history when the first generations of galaxies and quasars emitted radiation that ionized the intergalactic medium (IGM). Observations of the redshifted 21cm hyperfine transition of neutral hydrogen offer a novel 3D view of this era, encoding the distribution of ionized and neutral regions across cosmic time. Low-frequency radio interferometers such as the Hydrogen Epoch of Reionization Array (HERA), the Murchison Widefield Array (MWA), the LOw-Frequency ARray (LOFAR), and the Square Kilometre Array (SKA)\footnote{https://www.skao.int/en} are actively pursuing this signal through increasingly sensitive observations~\citep{Parsons_2010, Tingay_2013, DeBoer_2017, LOFAR, Mertens_2020, Trott_2020, Yoshiura_2021, Abdurashidova_2022, Abdurashidova_2023, Ceccotti_2025}. However, extracting robust astrophysical constraints from 21cm measurements remains extremely challenging due to the high dimensionality of the signal, non-Gaussianity, and the significant contamination from astrophysical foregrounds several orders of magnitude brighter than the cosmological signal.

Cross-correlations between 21cm observations and high-redshift galaxy surveys provide a promising avenue to both confirm the cosmological origin of the 21cm signal and enhance constraints on reionization models~\citep{Furlanetto2007, Park2014, Sobacchi2016, Vrbanec2016, Heneka2017, Hutter2017, Kubota2018, Yoshiura2018, Beane2019, Heneka2020, Vrbanec2020, Weinberger2020, Heneka2021, LaPlante2023, Moriwaki2024, Gagnon-Hartman2025, Hutter_2025}. Because foregrounds in radio interferometry are largely uncorrelated with galaxy positions, the 21cm galaxy cross-power spectrum is unbiased by foreground contamination~\citep{Furlanetto2007, Lidz2009, Yoshiura2018} and links the evolving ionization morphology to observable galaxy populations. Recent work has developed realistic forecasts for detecting the 21cm galaxy cross-power spectrum across instrument configurations and survey parameters, identifying regimes where signal-to-noise ratios (S/N) are maximized and where differing ionizing source scenarios can be distinguished observationally~\citep{Gagnon-Hartman2025, Hutter_2025}. These studies highlight how cross-power measurements can trace the evolving morphology of ionized regions and advance individual 21cm or galaxy-only probes of reionization. Upcoming facilities such as the Prime Focus Spectrograph (PFS) on Subaru~\citep{Greene2022}, the Nancy Grace Roman Space Telescope~\citep{Roman_2022}, the MOONRISE program~\citep[MOONS Extragalactic GTO;][]{Maiolino2020}, and the proposed Widefield Spectroscopic Telescope~\citep[WST;][]{WST_2024} will provide high-redshift galaxy samples spanning a range of survey areas, depths, and redshift precision, making practical survey optimization a pressing concern for maximizing scientific return from cross-correlation measurements.

With these diverse survey capabilities on the horizon, key questions remain regarding what astrophysical information cross-correlations can robustly extract and how survey design impacts scientific return. The neutral hydrogen fraction, $x_{\mathrm{HI}}(z),$ and the mean overdensity in neutral regions, $\langle 1+\delta_{\mathrm{HI}}\rangle(z),$ together encode both the EoR timeline and the ionization morphology (e.g., the distinction between reionization scenarios driven either by low-mass or massive galaxies)~\citep{Hutter2023b}. However, whether cross-power measurements can constrain these quantities under realistic noise and survey limitations remains uncertain. Furthermore, source properties such as the escape fraction of hydrogen-ionizing photons from galaxies and star formation efficiency---which fundamentally drive reionization---remain poorly constrained observationally, raising the question of whether cross-correlations offer a pathway to access these otherwise inaccessible parameters.

Addressing these questions requires robust statistical frameworks that can handle the intrinsic complexity of the joint data and the intractability of traditional likelihoods. The high-dimensional, non-Gaussian nature of the 21cm signal, combined with the absence of analytically accurate forward models relating astrophysical parameters to observables, renders traditional likelihood-based inference computationally prohibitive or impossible. Simulation-based inference (SBI), also known as likelihood-free inference, has recently emerged as a powerful tool in cosmological analyses~\citep{Alsing_2019, Cole_2022, Villaescusa-Navarro_2022, Saxena_2024, Schosser_2025, Ore_2025}, including the joint analysis of multiple probes~\citep{Schosser_2026}, circumventing explicit likelihood modeling by learning the mapping from observables to parameters directly from forward simulations. Our framework \texttt{EoRFlow}\footnote{https://github.com/astro-ML/EoRFlow}~\citep{Pietschke_2025}, originally trained on 21cm power spectra from diverse reionization models, exemplifies this approach. \texttt{EoRFlow} uses neural density estimators to efficiently and unbiasedly reconstruct the evolution of the global neutral hydrogen fraction $x_{\mathrm{HI}}(z)$ without relying on approximate likelihoods. This method enables scalable posterior estimation across narrow redshift slices and has been validated on realistic mock datasets incorporating instrument noise, demonstrating its potential for reconstructing reionization histories from forthcoming SKA-Low observations.

In this work, we extended the SBI paradigm to the 21cm galaxy cross-power spectrum, combining the realistic cross-power modeling developed in~\citet{Hutter_2025} with the flexible inference power of \texttt{EoRFlow}. By integrating simulated cross-power spectra with comprehensive noise and survey models into an SBI pipeline, we inferred the aforementioned astrophysical parameters governing reionization, directly linking the statistical imprint of high-redshift galaxies and neutral hydrogen fluctuations. This approach is essential because the absence of tractable analytic models connecting astrophysical parameters to cross-power observables makes SBI the natural framework for extracting physical information from these complex, high-dimensional measurements. Tackling cross-correlations with SBI not only enhances the robustness of parameter constraints by exploiting complementary information from the cross-power, it also mitigates the need for analytically tractable likelihoods in high-dimensional regimes where traditional methods struggle.

The structure of this paper is as follows. We begin with a description of our simulation framework and mock data generation in Section~\ref{sec: data}. Section~\ref{sec: network} then details the neural-network architecture and training procedures. Subsequently, Section~\ref{sec: results} presents our main results. We start with the inference of global reionization properties (Section~\ref{sec: EoR timeline}). We then systematically investigate the impact of galaxy survey parameters (Section~\ref{sec: galaxy survey}) and the information gained  from cross-correlation measurements (Section~\ref{sec: mutual info}). Moreover, we demonstrate the capabilities of the cross-power spectrum to infer reionization source properties (Section~\ref{sec: astro}). Finally, we conclude in Section~\ref{sec: conclusion}.

\section{The 21cm galaxy mock data}
\label{sec: data}

We generated mock realizations of the 21cm signal and the underlying galaxy population using \texttt{21cmFASTv4}~\citep{21cmfast11,21cmfast,Park_2019,Davies_2025}. The simulations produced coeval 3D cubes of halo properties and the corresponding differential 21cm brightness temperature, $\delta T_\mathrm{b}$, during the EoR.

We simulated coeval volumes at the redshifts $z\in[6.0, 6.3, 6.6, 7.0, 7.3, 7.6, 8.0]$, which span the late stages of reionization where current and forthcoming galaxy surveys, including PFS, Roman, MOONRISE, and WST, will enable the detection of 21cm galaxy cross-correlations through high-redshift galaxy samples. At these redshifts ($z\leq8$), we assumed a saturated spin temperature, as expected in standard EoR scenarios, corresponding to the post-heating regime in which $T_\mathrm{S} \gg T_\mathrm{CMB}$, and consistent with current upper limits from HERA~\citep{HERA2022, HERA2023, HERA2025}. Under this assumption, the 21cm brightness temperature depends only on the neutral hydrogen fraction and the density field via~\citep{Furlanetto2006}
\begin{align}\label{Tb}
    \delta T_\mathrm{b}(\mathbf{x}) \simeq T_0 x_\mathrm{HI}(\mathbf{x}) (1+\delta(\mathbf{x})).
\end{align}
All simulations were performed in a cubic volume of side length 300 cMpc with a cell size of 2 cMpc, providing sufficient resolution to capture the large-scale ionization morphology relevant for cross-correlation studies while maintaining computational efficiency.

To capture a broad range of physically plausible reionization scenarios, we varied the ionizing photon escape fraction, $f_\mathrm{esc}$, and the star formation efficiency, $f_*$, according to the power-law parameterization implemented in \texttt{21cmFASTv4}~\citep{Park_2019}:
\begin{align} \label{eq: astro params}
    f_*(M_\mathrm{h})&=f_{*,10} \left( \frac{M_\mathrm{h}}{10^{10}\mathrm{M}_\odot}\right)^{\alpha_*},\\
    f_\mathrm{esc}(M_\mathrm{h})&=f_{\mathrm{esc},10} \left( \frac{M_\mathrm{h}}{10^{10}\mathrm{M}_\odot}\right)^{\alpha_\mathrm{esc}},
\end{align}
where $f_{*,10}$ and $f_{\mathrm{esc},10}$, as well as $\alpha_*$ and $\alpha_\mathrm{esc}$, are the normalization and power-law exponents of the star formation efficiency and escape fraction, respectively.  We fixed other astrophysical and cosmological parameters to fiducial values consistent with Planck 2018 results~\citep[$\Omega_\mathrm{b}=0.049$, $\Omega_\mathrm{m}=0.31$, $h=0.68$;][]{Planck2018} and adopted the standard \texttt{21cmFAST} prescriptions for stellar spectra and X-ray heating. To exclude scenarios of extremely delayed or early reionization, we filtered out models where the neutral fraction remains either $x_{\mathrm{HI}}>0.7$ or $x_{\mathrm{HI}}<0.2$ across all redshifts in our observational window ($z=6.0$-$8.0$).
The simulation parameters and their prior ranges are summarized in Table~\ref{tab:parameters}. Uniform priors are assumed over the specified intervals.
\begin{table}[h]
    \caption{Summary of simulation parameters and prior ranges.}
    \label{tab:parameters}
    \centering
    \begin{tabular}{l c}
        \toprule
        \textbf{Parameter} & \textbf{Prior Range} \\
        \midrule
        $\log_{10}{f_\mathrm{esc,10}}$ & $\mathcal{U}[0.005, 0.5]$ \\
        $\alpha_\mathrm{esc}$ & $\mathcal{U}[-0.8, 0.5]$\\
        $\log_{10}{f_{*,10}}$ & $\mathcal{U}[0.005, 0.5]$\\
        $\alpha_*$ & $\mathcal{U}[-0.3,0.9]$\\
        \bottomrule
    \end{tabular}

\end{table}
These parameters jointly control the abundance and clustering of ionizing sources and therefore strongly influence both the morphology of ionized regions and the amplitude and scale dependence of the 21cm galaxy cross-power spectrum. In addition to these astrophysical parameters, we also varied the random seed for each simulation to account for cosmic variance.

To create realistic mock observations of the 21cm auto-power spectrum and the 21cm galaxy cross-power spectrum, we employed the forward-modeling framework developed in~\citet{Hutter_2025}. This method incorporates instrumental noise, survey geometry, and galaxy-selection effects in a self-consistent manner, enabling direct comparison with realistic observational scenarios.
From each simulated coeval cube, we computed the spherically averaged 1D power spectra $P_{21}(k)$ and $P_\mathrm{21,g}(k)$. For the galaxy field, halos were selected according to survey-specific detection thresholds, and their spatial distribution was used to construct the galaxy overdensity field.
The cross-power spectrum is then computed by correlating the galaxy overdensity with the 21cm brightness temperature fluctuations. Here, we used 15 linearly spaced bins with $k\in [0.08,2.40]$. 
Thermal noise contributions to the 21cm measurements and shot noise in the galaxy field were added following the prescriptions described in~\citet{Hutter_2025}. 
The 21cm noise calculations are based on the public code \texttt{21cmSense}~\citep{21cmsense13, 21cmsense14}. We assumed 100h of SKA AA* observations as well as the moderate foreground model in which the foreground wedge extends $0.1\mathrm{Mpc}^{-1}$ beyond the horizon limit. For the fiducial galaxy survey configuration used in Section~\ref{sec: EoR timeline}, we adopted a field of view (FOV) of $100\mathrm{deg}^2$, spectroscopic redshift precision $\sigma_z=0.001$, and a minimum halo mass of $M_{\mathrm{h,min}}=10^{11}M_\odot$, representing an optimistic but realistic large-area spectroscopic survey, as proposed for the WST. In Section~\ref{sec: galaxy survey} we systematically explore the impact of varying these parameters. For the largest FOV considered here ($100\mathrm{deg}^2$), the signal power is computed from the simulation box (side length 300cMpc), while the uncertainties are scaled to the larger FOV by accounting for the increased number of k-modes.

Note that in this work we treated halos and their attributed masses as proxies for galaxies. For cross-correlation studies, emission-line galaxy surveys---such as Lyman-$\alpha$ and [OIII] emitters---are ideal due to their precise redshift measurements. To the first order, the luminosities of these emission lines scale with the star formation rate, which increases with halo mass, justifying our use of halo mass as a proxy for galaxy detectability. Accordingly, halo masses can be translated to their corresponding line luminosities through empirical models. For example, halo masses of $M_\mathrm{h} = 10^{10}$, $10^{10.5}$, and $10^{11}\mathrm{M}_\odot$ correspond to Lyman-$\alpha$ luminosities of $L_{\mathrm{Ly}\alpha} \approx 10^{41}$, $10^{41.8}$, and $10^{42.3}\mathrm{erg\,s^{-1}}$, respectively, at $z \sim 7$ \citep[see Fig. 5 in][]{Hutter2023a}.

\begin{figure}
    \centering
    \includegraphics[width=0.49\textwidth]{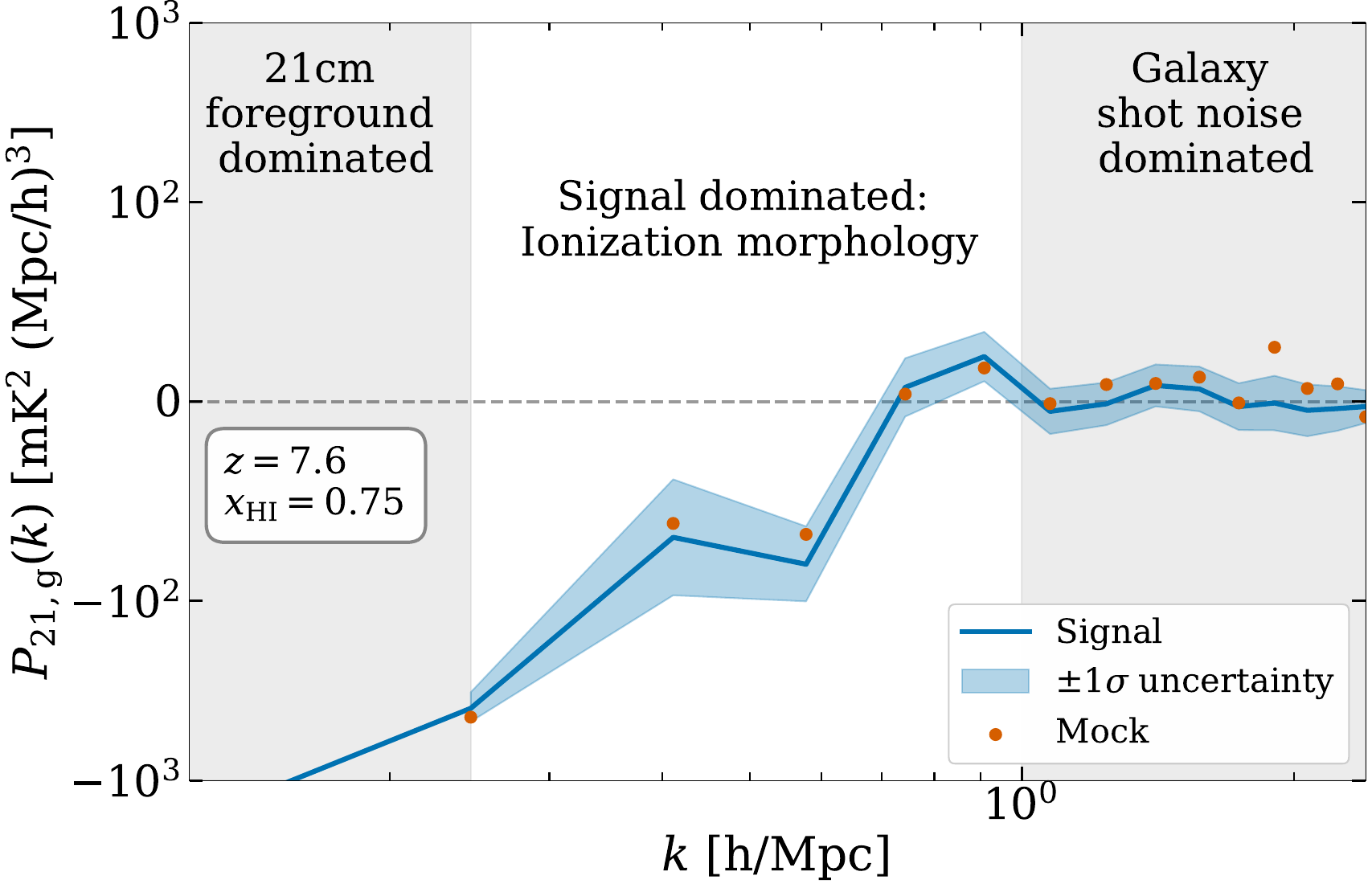}
    \caption{
    The 21cm galaxy cross-power spectrum at $z=7.6$ ($x_{\mathrm{HI}}=0.75$) for the fiducial survey configuration explored in the inference analysis in Section~\ref{sec: EoR timeline} (${\mathrm{FOV}}=100 {\mathrm{deg}}^2$, $\sigma_z=0.001$, $M_{\mathrm{ h,min}}=10^{11} M_\odot$). The blue line shows the physical signal, the shaded blue region indicates the $1\sigma$ uncertainty, and orange points represent the mock observation including instrumental noise. Three distinct regimes characterize the cross-power spectrum. The first region corresponds to the 21cm foreground wedge where large-scale modes are contaminated by bright foreground emission, resulting in poor S/N. We omitted the signal here based on our foreground-avoidance strategy. The second region is signal dominated and provides access to the ionization morphology. It exhibits strong anticorrelation on intermediate scales ($k\sim0.25$-$0.75 h {\mathrm{Mpc}}^{-1}$) where galaxies predominantly occupy ionized bubbles, while the 21cm signal traces neutral hydrogen in the IGM. At $k\sim0.75 h {\mathrm{Mpc}}^{-1}$, the zero crossing corresponds to the characteristic scale of ionized regions. On scales smaller than the typical size of ionized bubbles, both the galaxy distribution and neutral hydrogen density respond to the same underlying matter fluctuations, resulting in positive correlation. The third region shows the shot-noise-dominated regime on small scales ($k>1.0 h {\mathrm{Mpc}}^{-1}$) owing to the discrete nature of galaxy datasets.
    }
    \label{fig:cross power example}
\end{figure}

To generate mock observations of the power spectra that reflect the combined impact of astrophysical modeling and observational limitations, we sampled from a normal distribution of $\mathcal{N}(P_\mathrm{21,g}(k), \sigma_\mathrm{21,g}(k))$, where $P_\mathrm{21,g}(k)$ is the true, physical cross-power spectrum and $\sigma_\mathrm{21,g}(k)$ represents the observational uncertainties determined by the FOV; redshift precision, $\sigma_z$; and minimum halo mass, $M_\mathrm{h,min}$; as well as thermal noise from SKA-Low AA* (100h) and foreground avoidance. An example cross-power spectrum is shown in Fig.~\ref{fig:cross power example}.
In total, we produced 10790 samples (after filtering) with varying astrophysical parameters and random seed. 

One of our goals is to infer the full posterior probability distribution of neutral hydrogen fractions $x_\mathrm{HI}(z)$ and the mean overdensity in neutral regions, $\langle 1+\delta_\mathrm{HI} \rangle (z)$, at different redshifts, $z$ from this mock data. The latter can be obtained from Eq.\eqref{Tb}. At a given redshift we computed the mean differential brightness temperature, $\langle \delta T_\mathrm{b} \rangle,$ and neutral fraction, $\bar x_\mathrm{HI} $. For a standard cosmology with $\Omega_\mathrm{b}=0.049, \Omega_\mathrm{m}=0.31$ and $h=0.68$, the normalization factor is given by $T_0 = 27 \sqrt{0.1 (1+z)}\, \mathrm{mK}$~\citep{21cmfast11}, from which the average density can then be calculated as
\begin{align}
    \langle 1+\delta_\mathrm{HI}\rangle(z) = \frac{\langle \delta T_\mathrm{b}\rangle (z)} {27 \sqrt{0.1 (1+z)} \bar x_\mathrm{HI} (z)\, \mathrm{mK}}.  
\end{align}
Here, $\bar{x}_\mathrm{HI}(z)$ denotes the volume-averaged neutral fraction, 
and $\langle 1+\delta_\mathrm{HI}\rangle(z)$ is the mean density contrast 
averaged over neutral cells only. This separation is exact in \texttt{21cmFAST}, 
where $x_\mathrm{HI}(\mathbf{x})$ is binary for each cell.
These quantities are directly computed from the 21cm simulation outputs and saved as labels for network training.
This parameter combination enables constraints on both the EoR timeline and the distinction of different reionization scenarios~\citep{Hutter2023b}.

\section{Network architecture and training}
\label{sec: network}

We performed SBI using \texttt{EoRFlow}, a conditional normalizing flow framework originally introduced in~\citet{Pietschke_2025} for inference on 21cm power spectra.
We also explored an inference model based on conditional flow matching~\citep[CFM;][]{Lipman2023, Tong2023}, motivated by recent successes in modeling complex distributions with more flexible continuous-time dynamics. While CFM achieves comparable posterior accuracy, posterior sampling is significantly slower (about 6x in our case) due to the need to numerically integrate an ordinary differential equation for each draw. Given the absence of clear performance gains and the substantially increased computational cost, we do not adopt CFM as our fiducial inference model.

In the present work, \texttt{EoRFlow} was adapted to operate on 21cm galaxy cross-power spectra, while preserving the core design philosophy of directly inferring astrophysical parameters from summary statistics derived from forward simulations, without constructing an explicit likelihood.
The inference network directly took the power spectrum as input, and no additional neural summary network was employed. This choice reflects the relatively low dimensionality of the spherically averaged power spectrum and avoids introducing unnecessary architectural complexity or information bottlenecks. The network input consists of spherically averaged 1D power spectra; the 21cm auto-power, $P_{21}(k, z)$; the 21cm galaxy cross-power, $P_\mathrm{21,g}(k, z)$; or both combined, evaluated in discrete redshift bins. This flexibility to seamlessly incorporate single or multiple observables is a key advantage of the SBI approach. For each redshift slice, the power spectrum is sampled on a fixed set of k-modes, forming a 1D vector that fully characterizes the observable(s) used for inference.

The inference network is implemented as a conditional normalizing flow, which models the posterior distribution $p(\theta | \mathbf{x})$, where $\theta$ denotes the astrophysical parameters of interest and $\mathbf{x}$ the input cross-power spectrum.
The overall flow structure follows~\citet{Pietschke_2025} and consists of a sequence of coupling transformations interleaved with permutations to ensure full mixing of parameters. However, affine coupling layers were replaced by rational quadratic spline coupling layers~\citep{Durkan_Spline}.
Spline-based coupling layers provide greater expressivity than affine transformations, allowing the flow to model highly non-Gaussian posterior distributions with fewer layers. This increased flexibility is particularly important for 21cm galaxy cross-correlation data, where the posterior structure can be influenced by complex interactions between galaxy bias, ionization morphology, and observational noise.
We employed ten coupling layers with 512 nodes each and ReLU activation. 
All training and evaluation was performed in PyTorch~\citep{Pytorch} with the AdamW optimizer and default decoupled weight decay for regularization~\citep{AdamW}.

From the entire dataset of 10790 samples, we used 7832 for training, 1958 for validation, and 1000 for testing. All evaluations shown in this work were performed on the test set. 
For each sample the network condition consists of one power spectrum per redshift, each of which consists of 15 k-bins. For our seven redshifts this results in a total flattened array size of 105. We also appended the redshift values for temporal context, bringing the total condition dimension to 112. 
The power spectrum values were preprocessed by an inverse hyperbolic sine (asinh) transformation, which handles the wide dynamic range (similar to a logarithmic transformation) while accepting both positive and negative values, as present in the cross-power spectrum. Subsequently, we applied z-score normalization. Additionally, we transformed the $x_\mathrm{HI}$ labels using a logit transformation. 
We trained the model with a batch size of 16 and an initial learning rate of $10^{-3}$, which was halved every ten epochs without improvement in the validation loss. Early stopping was employed to avoid overfitting.
Our fiducial model, trained on the cross-power spectrum for the survey configuration described in Section~\ref{sec: data}, converges after 40 epochs. When including the 21cm auto-power spectrum alongside the cross-power, training lasted for 54 epochs. On one NVIDIA A100 GPU, training took approximately 20--30 minutes, depending on survey parameters. 

\section{Results} \label{sec: results}

In this section, we demonstrate the inference capabilities of our SBI framework by applying the trained \texttt{EoRFlow} to mock power-spectrum observations spanning a range of observational scenarios. Throughout this analysis, we focused on 1D power spectra. Even though the 2D cross-power spectrum should, in principle, be more sensitive to the ionization morphology and galaxy bias, we found that the higher S/N through spherical averaging surpasses this information loss. A comparison of the impact of power spectrum dimensionality can be found in Appendix~\ref{app: dimension}.

We organized our results into four complementary analyses. Assuming a fiducial survey configuration (see Section~\ref{sec: data}), we began by constraining the neutral hydrogen fraction, $x_\mathrm{HI}(z),$ and the mean density contrast in neutral regions $\langle 1+\delta_\mathrm{HI} \rangle(z)$ across seven redshift slices (Section~\ref{sec: EoR timeline}), comparing three training configurations: 21cm auto-power alone, cross-power alone, and both combined. For the combined model, we concatenated independently normalized power spectra into a single conditioning vector, doubling its dimensionality. While this in principle could benefit from architectural adjustments to optimally exploit the higher dimensional input, we deliberately fixed the network architecture to be identical across all three configurations to ensure fair comparison. Any performance differences thus reflect the information content of the data rather than hyperparameter tuning.

Once the baseline inference performance for our fiducial survey is established, we systematically varied galaxy survey parameters (FOV, $\sigma_z$, $M_\mathrm{h,min}$) to understand which observational characteristics most strongly influence inference quality and to identify survey designs that optimize scientific return (Section~\ref{sec: galaxy survey}). Afterwards, we employed a mutual information analysis to decompose the complementary information between auto-power and cross-power measurements on a per-bin basis, revealing which scales and redshifts benefit most from multi-tracer observations (Section~\ref{sec: mutual info}). Finally, we extended our analysis beyond global quantities to infer the astrophysical properties of ionizing sources, demonstrating that cross-correlation measurements provide access to parameters that are fundamentally inaccessible to 21cm auto-power observations alone (Section~\ref{sec: astro}).

\subsection{Inferring global reionization properties} \label{sec: EoR timeline}
\begin{figure*}[h]
    \centering
    \includegraphics[width=0.85\textwidth]{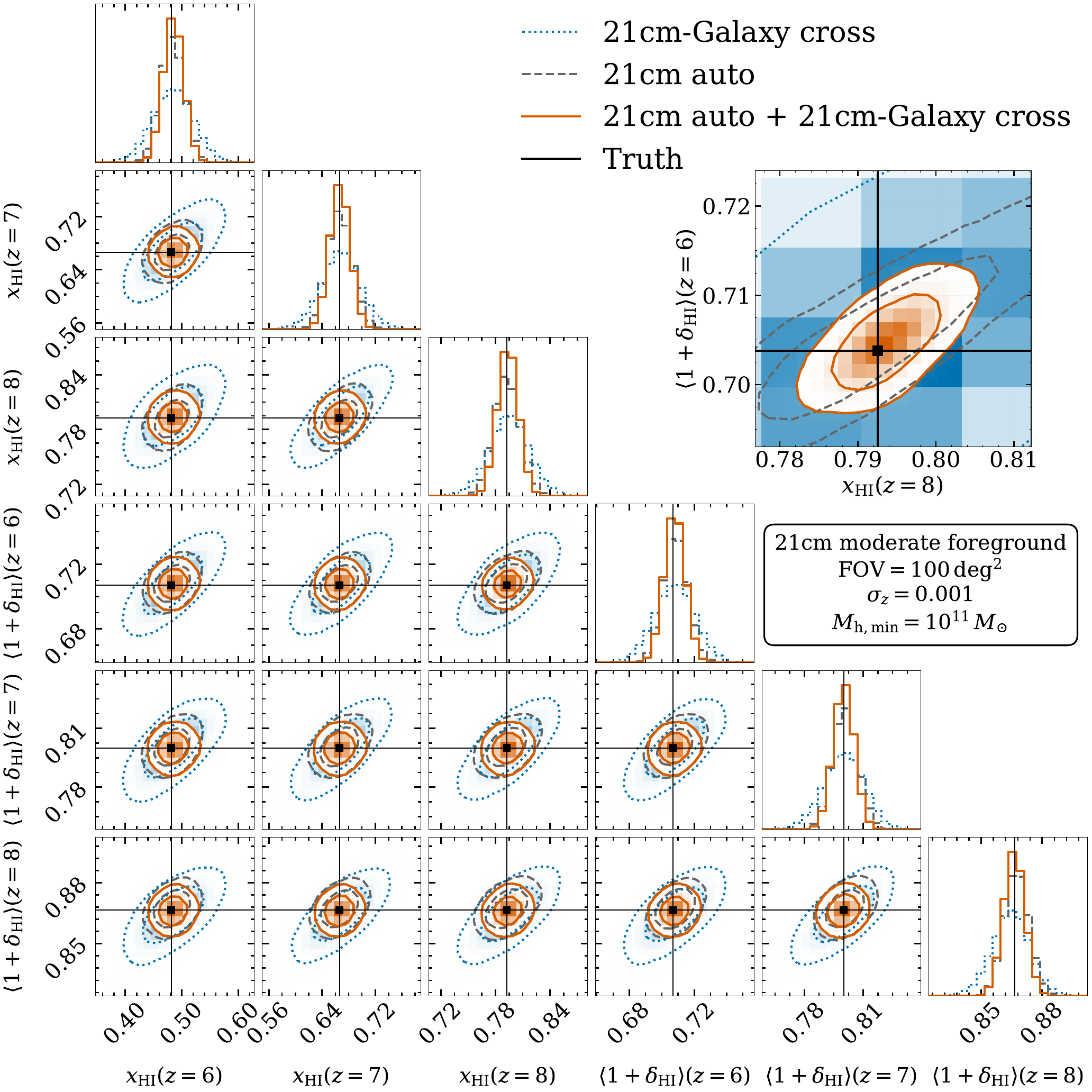}
    \caption{Marginalized posteriors for neutral fraction, $x_\mathrm{HI}(z),$ and the mean density in neutral regions, $\langle 1+\delta_\mathrm{HI} \rangle (z),$ for a selection of redshift slices and a randomly chosen set of parameters ($\log_{10} f_{*,10}=0.04$,  $\log_{10} f_{\mathrm{esc}, 10}=0.05$, $\alpha_*=0.44,$ and $\alpha_\mathrm{esc}=-0.60$) assuming moderate 21cm foreground avoidance. Shadings indicate the 68\% and 95\% confidence intervals. Dotted blue contours show the posteriors derived from 21cm galaxy cross-power spectra, while solid orange lines represent the results when combining both cross- and auto-power spectra. The dashed gray contours denote the model trained on 21cm auto-power spectra only, and true parameter values are indicated by black dots. This realization serves as an illustrative example; posterior calibration was confirmed via coverage statistics following~\citet{Pietschke_2025}.}
    \label{fig:corner}
\end{figure*}
We begin by demonstrating the inference capabilities of our framework for constraining the global evolution of reionization, specifically the neutral hydrogen fraction, $x_\mathrm{HI}(z),$ and the mean density contrast in neutral regions, $\langle 1+\delta_\mathrm{HI} \rangle(z),$ across seven redshift slices spanning $z = 6.0$ to $8.0$. These quantities provide complementary views of the reionization process. The neutral fraction directly quantifies the progression of reionization, while the mean density in neutral regions reflects the connection between the matter distribution and the ionization morphology, capturing how reionization proceeds preferentially in overdense regions where the first galaxies formed.

For this analysis, we adopted a fiducial observational configuration that combines moderate foreground assumptions with realistic survey parameters as detailed in Section~\ref{sec: data}. This configuration represents an ambitious but achievable observational program for the coming decade, combining SKA-Low 21cm observations with wide-field spectroscopic galaxy surveys such as the WST, and it serves as our baseline for comparing different data combinations.
We trained three separate conditional normalizing flow models on mock observations drawn from this fiducial survey, using power spectra from 1000 test realizations to evaluate inference performance. As described above, the three models differ only in their input data: using the 21cm auto-power spectrum alone, the cross-power spectrum alone, or both combined.

Figure~\ref{fig:corner} presents the marginalized posterior distributions for a representative test observation with randomly chosen parameters: $\log_{10} f_{*,10}=0.04$,  $\log_{10} f_{\mathrm{esc}, 10}=0.05$, $\alpha_*=0.44,$ and $\alpha_\mathrm{esc}=-0.60$. The corner plot shows the inferred neutral fraction and mean density across four selected redshift slices. The 68\% and 95\% confidence regions reveal several key features of the inference quality. First of all, all three models recover the true parameter values (marked by black dots) within their respective uncertainty intervals for this particular realization, which was selected as a representative illustration. The calibration of our inference method was verified through coverage statistics following \citet{Pietschke_2025}.
\begin{figure*}
\sidecaption
    \centering
    \includegraphics[width=12cm]{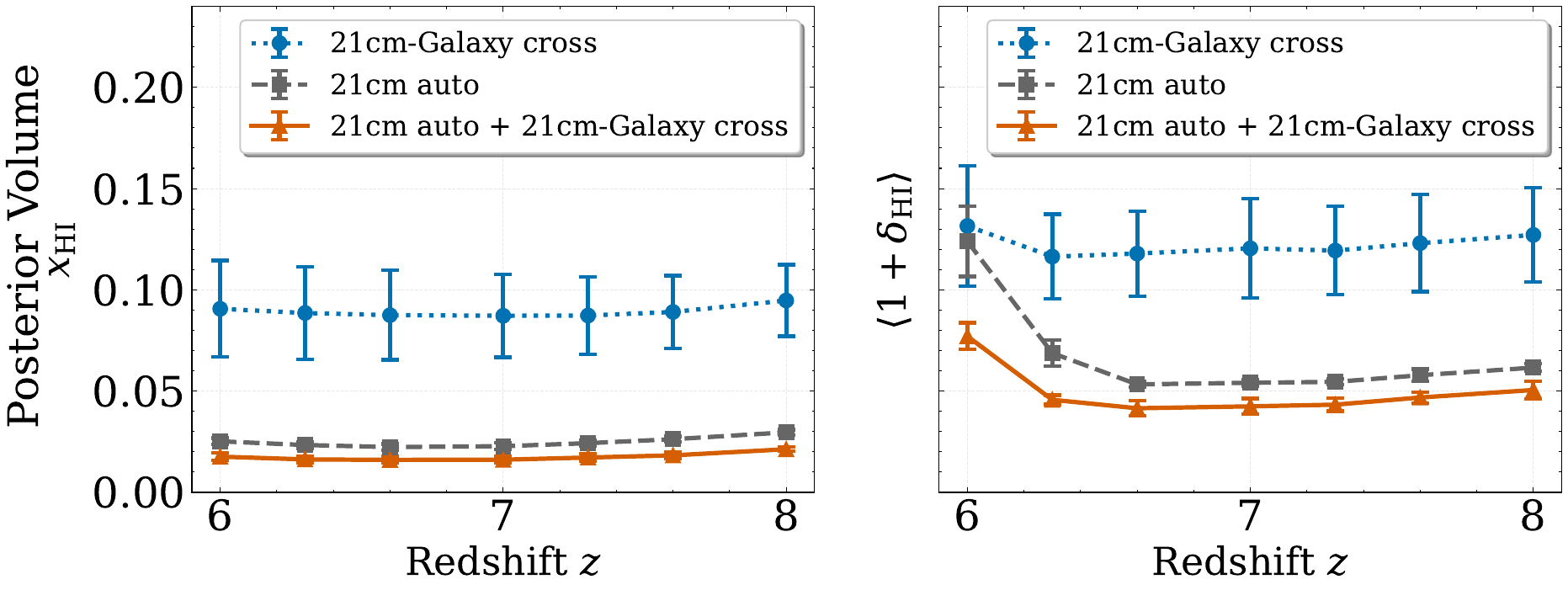}
    \caption{Informativeness measured by the normalized posterior volume on the 
    neutral fraction $x_\mathrm{HI}(z)$ and the mean density in neutral regions 
    $\langle 1+\delta_\mathrm{HI} \rangle (z)$ as a function of redshift assuming 
    moderate 21cm foreground avoidance as well as our fiducial galaxy survey 
    (100 deg$^2$, $\sigma_z = 0.001$, $M_{\mathrm{h,min}} = 10^{11}$ M$_\odot$). 
    Points show the mean posterior volume across 6 independent network training runs and error bars indicate the standard deviation.}
    \label{fig: posterior volume}
\end{figure*}
Moreover, the posteriors for $x_\mathrm{HI}$ are generally somewhat more tightly constrained than those for $\langle 1+\delta_\mathrm{HI} \rangle$, reflecting the fact that the neutral fraction is the primary driver of power-spectrum amplitude, while density fluctuations contribute a secondary modulation. Furthermore, the relative sizes of the confidence regions immediately reveal the hierarchy of constraining power among the three data combinations, with the combined model (solid orange contours) providing the tightest constraints, followed by the 21cm auto-power model (dashed gray contours), and finally the cross-power-only model (dotted blue contours).

To quantify this hierarchy more precisely across the full test set, we computed the posterior volume (PV) for each model, which is defined as the standard deviation of the posterior samples normalized by the prior standard deviation. This metric measures what fraction of the prior uncertainty remains after observing the data, with values near unity indicating uninformative observations and values near zero indicating precise constraints. Fig.~\ref{fig: posterior volume} shows the mean PV across six independent network training runs as a function of redshift for both $x_\mathrm{HI}$ (left panel) and $\langle 1+\delta_\mathrm{HI} \rangle$ (right panel), with error bars indicating the standard deviation. Averaging over multiple runs accounts for the epistemic uncertainty inherent to neural-network training, where different random initializations can yield slightly different posteriors; this is discussed in detail in Appendix~\ref{app: stochasticity}. 

The results confirm the qualitative impression from the corner plot while revealing additional structure. For the neutral fraction, the 21cm auto-power spectrum (gray) achieves remarkably tight constraints with the PV below 0.03 across all redshifts, demonstrating the direct sensitivity of the 21cm signal to $x_\mathrm{HI}$. The cross-power spectrum alone (blue) yields broader posteriors with a PV around 0.09, reflecting its indirect connection to the neutral fraction through the correlation between ionization bubbles and galaxies. Combining both measurements (orange) further improves constraints to the PV below 0.02, indicating that despite the strong baseline from auto-power alone, the cross-power spectrum does provide modest complementary information. 

For the mean neutral hydrogen-density contrast, the pattern is similar but generally has a larger PV, particularly at lower redshifts. This redshift dependence likely reflects the decreasing size and contrast of neutral regions as reionization progresses toward completion at $z \sim 6$, making the density signal weaker and more challenging to extract. Across all redshifts, the combined model achieves a PV around 0.04, auto-power alone reaches 0.07, and cross-power alone yields 0.12. In both cases, the improvements from combining measurements are statistically significant but modest, typically yielding 20--30\% reductions in posterior width compared to auto-power alone.

These findings establish several important conclusions for interpreting the remainder of our results. Firstly, 21cm auto-power spectra alone provide excellent constraints on the global reionization history, demonstrating the power of direct neutral hydrogen observations, which confirms the findings of previous works~\citep{Pietschke_2025, Cooper_2025, Cerardi_2025}. 
Furthermore, cross-power measurements do contain useful complementary information about both the ionization state and the density field, though this information is secondary to what auto-power spectra provide for these particular parameters. The modest improvements seen here, however, should not be interpreted as evidence that cross-power measurements are of limited scientific value. As we demonstrate in subsequent sections, the information concerning cross-power spectra becomes dramatically more important when considering parameters that are poorly constrained by auto-power shape alone, particularly those related to the properties of ionizing sources. The hierarchy observed here (both better than "auto", which is much better than "cross") reflects the fact that global neutral fraction and density are most directly encoded in the 21cm signal itself, while cross-correlations shine when probing the spatial relationships between ionization and structure formation.

\subsection{Impact of galaxy survey parameters} \label{sec: galaxy survey}
Having established baseline inference performance for our fiducial survey configuration, we now investigate how variations in galaxy-survey design affect the constraining power of 21cm galaxy cross-power measurements. This analysis addresses a practical question faced by the community as to which observational parameters most critically determine scientific return, and which design choices offer the greatest leverage for optimizing future surveys. Understanding these dependencies is essential for guiding observational strategies and evaluating trade-offs among survey area, depth, and redshift precision.
\begin{figure*}[h!]
    \centering
    \includegraphics[width=0.9\textwidth]{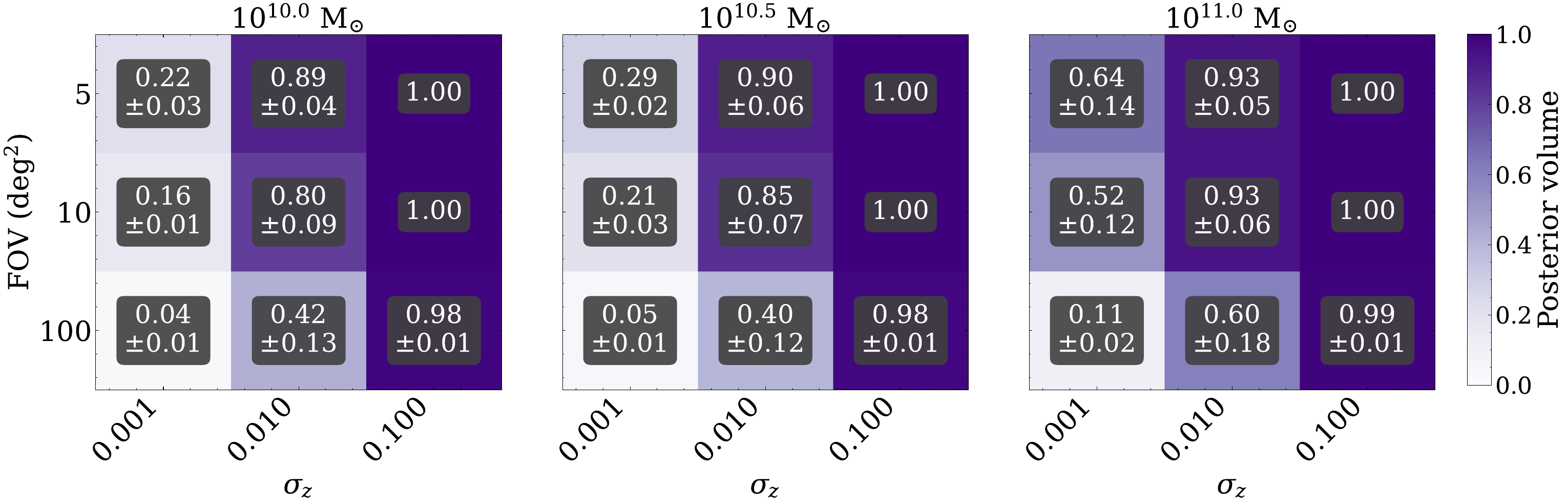}
    \caption{Posterior volume (averaged over neutral fraction, density, redshifts, and test observations) as a function of galaxy survey parameters assuming moderate 21cm foreground avoidance. Each panel shows results for a different minimum detectable halo mass threshold. The color-scale ranges from white (posterior volume near zero, highly constraining) to dark purple (posterior volume near unity, uninformative), with numerical values annotated in each cell. The left panel shows $M_\mathrm{h,min} = 10^{10}$ M$_\odot$, the center panel $10^{10.5}$ M$_\odot$, and the right panel $10^{11}$ M$_\odot$. The FOV increases from top to bottom, and redshift precision improves from right to left. Each cell shows the mean and standard deviations of the PV computed over six independent network training runs; the epistemic uncertainty is discussed in Appendix~\ref{app: stochasticity}.}
    \label{fig:surveys}
\end{figure*}
We kept the 21cm observational component fixed throughout this analysis, assuming 100h of SKA-Low observations using the AA* array configuration with the moderate foreground avoidance model, as already detailed in Section~\ref{sec: EoR timeline}.
We then systematically varied three main galaxy survey parameters, each spanning a physically motivated range that encompasses both conservative assumptions and optimistic future capabilities. The FOV spans $\{5, 10, 100\}$ deg$^2$, representing various wide-area surveys. The redshift precision varies as $\sigma_z \in \{0.001, 0.01, 0.1\}$, corresponding to spectroscopy, grism, and photometric redshifts. The minimum detectable halo mass threshold takes values of $M_\mathrm{h,min} \in \{10^{10}, 10^{10.5}, 10^{11}\}$ M$_\odot$, representing extremely deep surveys, deep surveys, and standard galaxy surveys, respectively. These parameter ranges, summarized in Table~\ref{tab:survey parameters}, encompass several forthcoming high-redshift galaxy surveys.
\begin{table}[h]
    \centering
    \caption{Galaxy survey parameters explored in this section.}
    \begin{tabular}{l c}
        \toprule
        \textbf{Parameter} & \textbf{Values} \\
        \midrule
        $\mathrm{FOV} [\mathrm{deg}^2]$ & $[5,10,100]$ \\
        $\sigma_z$ & $[0.001, 0.01, 0.1]$\\
        $M_\mathrm{h,min} [\mathrm{M}_\odot]$ & $[10^{10}, 10^{10.5}, 10^{11}]$\\
        \bottomrule
    \end{tabular}
    \tablefoot{All $3^3 = 27$ combinations were evaluated while keeping 21cm observational parameters fixed at the fiducial configuration (100h SKA AA* with moderate foreground avoidance).}
    \label{tab:survey parameters}
\end{table}
Spectroscopic configurations ($\sigma_z = 0.001$) correspond to facilities such as the PFS on Subaru, which can achieve intermediate FOVs ($\sim 10$ deg$^2$) with excellent redshift precision, the proposed WST, which would enable wide-area spectroscopic surveys approaching 100 deg$^2$ and the MOONRISE program providing spectroscopic observations over $\sim 1$ deg$^2$. Grism observations ($\sigma_z = 0.01$) are within reach of the Nancy Grace Roman Space Telescope, which is expected to survey 5--10 deg$^2$ in grism mode.  Photometric-like precision ($\sigma_z = 0.1$) represents either multiband narrow-band surveys or prism observations. Our fiducial configuration (100 deg$^2$, $\sigma_z = 0.001$, $M_{\mathrm{h,min}} = 10^{11}$ M$_\odot$) represents an optimistic but feasible large-area spectroscopic program comparable to an extended PFS campaign or future WST observations, while more conservative, near-term surveys occupy the intermediate parameter space with smaller FOVs and a potentially coarser redshift precision.

Evaluating all $3^3 = 27$ parameter combinations requires training separate \texttt{EoRFlow} models for each survey configuration, as the noise properties and signal characteristics vary substantially across this parameter space. Thanks to the computational efficiency of our framework, this comprehensive survey is feasible. Training a single model requires up to 30 minutes on a single Nvidia A100 GPU depending on survey parameters. To additionally quantify epistemic uncertainty from network training stochasticity, each configuration was trained six times with different random initializations, amounting to 162 models in total. The total computational cost amounts to approximately 70 hours on one GPU, demonstrating the scalability of SBI for survey-optimization studies that would be prohibitive with traditional inference approaches. After training, we evaluated each model on its respective held-out test set of 1000 observations to ensure fair comparison.
For each survey configuration, we computed the PV metric defined in Section~\ref{sec: EoR timeline}, averaged over both $x_\mathrm{HI}(z)$ and $\langle 1+\delta_\mathrm{HI} \rangle(z)$ across all redshift slices and all test observations. This single number summarizes the overall constraining power of each survey design, with values near zero indicating tight constraints and values near unity indicating uninformative observations. Fig.~\ref{fig:surveys} presents these results as a three-panel heat map, with each panel corresponding to a fixed halo mass threshold and showing PV as a function of FOV and redshift precision.

Several clear trends emerge from this parameter exploration. First and most strikingly, redshift precision dominates the constraining power across all configurations. Surveys with poor redshift accuracy ($\sigma_z = 0.1$, corresponding to photometric estimates) yield a PV near unity regardless of FOV or depth, indicating they provide essentially no useful constraints on reionization parameters. Even grism quality ($\sigma_z = 0.01$) shows substantially degraded performance compared to spectroscopy ($\sigma_z = 0.001$). This strong dependence arises because the constraining power of the cross-power spectrum relies on accurate line-of-sight mode measurements. Redshift uncertainties suppress the detection of the line-of-sight component of all Fourier modes, with the effect being most severe for modes corresponding to spatial scales smaller than those probed by $\sigma_z$.

Furthermore, the interplay between FOV and halo-mass threshold reveals important survey design considerations. For the most conservative halo-mass limit ($M_\mathrm{h,min} = 10^{11}$ M$_\odot$, right panel), only the widest survey (100 deg$^2$) combined with excellent redshift precision ($\sigma_z = 0.001$) achieves good constraints (PV $\sim 0.1$). This configuration corresponds to our fiducial WST-type survey from Section~\ref{sec: EoR timeline}. All other parameter combinations at this halo-mass threshold yield a PV of $\gtrsim 0.5$. While configurations with $\sigma_z = 0.01$ and a large FOV can reduce the prior volume by approximately 50\%, this remains modest compared to the spectroscopic case. The physical interpretation is straightforward. At $M_\mathrm{h,min} = 10^{11}$ M$_\odot$, galaxy number densities are relatively low, requiring wide areas to accumulate sufficient statistics for measuring the cross-power signal above noise.

Moving to deeper surveys dramatically relaxes the FOV requirement. At $M_\mathrm{h,min} = 10^{10.5}$ M$_\odot$ (center panel), even the 10 deg$^2$ field achieves a PV of around 0.2 when combined with $\sigma_z = 0.001$, representing good constraints. At the most optimistic depth of $M_\mathrm{h,min} = 10^{10}$ M$_\odot$ (left panel), all field sizes yield excellent constraints (PV $\lesssim 0.2$) provided redshift precision remains high. This trend reflects the rapid increase in galaxy number density toward lower mass thresholds, which enhances the statistical precision of the cross-power measurement through increased mode sampling. Deep surveys thus offer a path to high-quality science with more modest survey areas, though achieving such depths remains observationally challenging.

Intermediate redshift precision shows more modest gains. The central column of each panel shows that $\sigma_z = 0.01$ configurations 
provide intermediate constraints, though these are substantially degraded compared to 
spectroscopic precision ($\sigma_z = 0.001$). We caution, however, that these 
intermediate configurations carry the largest epistemic uncertainty from network 
training stochasticity (see Appendix~\ref{app: stochasticity}), and their PV 
values should be interpreted qualitatively rather than as precise quantitative 
measurements. The overall finding is nonetheless clear: survey strategies should 
prioritize redshift precision over area or depth when trade-offs are necessary. 
Investing in high-resolution spectroscopy ($\sigma_z \sim 0.001$) is essential 
for extracting meaningful constraints from cross-power measurements. Once 
sufficient galaxy number density is achieved, further improvements from expanding 
survey area or lowering mass thresholds provide diminishing returns compared to 
maintaining redshift accuracy.

These results carry important implications for planning future high-redshift galaxy surveys intended for cross-correlation studies with 21cm experiments. Our fiducial configuration (100 deg$^2$, $\sigma_z = 0.001$, $M_\mathrm{h,min} = 10^{11}$ M$_\odot$) represents an ambitious but realistic target for wide spectroscopic surveys that would be possible with the WST. However, the analysis demonstrates that substantial scientific gains are possible with less extreme survey areas if deeper observations can reach $M_\mathrm{h,min} \sim 10^{10.5}$ M$_\odot$ while maintaining excellent redshift precision. Conversely, photometric or grism surveys will struggle to provide competitive constraints regardless of area or depth, highlighting the critical importance of redshift accuracy for this science case.

Finally, we emphasize that these conclusions are specific to the inference targets considered here, namely the neutral fraction and mean density evolution. As demonstrated in Section~\ref{sec: astro}, different astrophysical parameters may exhibit different sensitivities to survey design, with source properties requiring deeper observations. The survey optimization landscape thus depends critically on the scientific questions being addressed, and comprehensive survey design must consider the full range of parameters of interest rather than optimizing for a single metric.

\subsection{Information gained from cross-correlation measurements} \label{sec: mutual info}
While the previous sections examined the constraining power of 21cm auto-power and 21cm galaxy cross-power spectra, it is valuable to quantify the per-bin information gained by including the cross-correlation measurements. Specifically, given that 21cm observations provide constraints in each redshift, $z,$ and wave-number, $k,$ bin, we wanted to know how much additional information the cross-power spectrum provides for each $(z,k)$ pair.
To address this question, we employed an information--theoretical approach based on mutual information~\citep[MI;][]{Cover2006}. MI quantifies the reduction in uncertainty about one random variable when another is observed. For discrete random variables $X$ and $Y$, it is defined as
\begin{equation}
    I(X; Y) = \sum_{x,y} p(x,y) \log \frac{p(x,y)}{p(x)p(y)}, \label{eq:MI}
\end{equation}
where $p(x,y)$ is the joint probability distribution and $p(x)$, $p(y)$ are the marginal distributions. Conditional MI $I(X;Y|Z)$ measures the information gained about $X$ from observing $Y$ when $Z$ is already known, and it can be expressed as
\begin{equation}
    I(X;Y|Z) = I(X; [Y,Z]) - I(X;Z), \label{eq: CMI}
\end{equation}
where $I(X; [Y,Z])$ is the MI between $X$ and the joint observation of both $Y$ and $Z$.

We computed the conditional MI for each parameter, $\theta_j$ (where $j$ indexes either the neutral fraction, $x_\mathrm{HI}(z),$ or the mean density contrast, $\langle 1+\delta_\mathrm{HI} \rangle(z)$), with respect to the cross-power spectrum conditioned on the auto-power spectrum for each $(z,k)$ bin:
\begin{equation}
    I(\theta_j; \mathrm{cross}_{z,k} | \mathrm{auto}_{z,k}) = I(\theta_j; [\mathrm{auto}_{z,k}, \mathrm{cross}_{z,k}]) - I(\theta_j; \mathrm{auto}_{z,k}).
\end{equation}
This quantity represents the additional information about the parameter $\theta_j$ provided by the cross-power at the bin $(z,k)$ beyond what the auto-power already provides. We further computed the fractional information gained as the ratio $I(\theta_j; \mathrm{cross}_{z,k} | \mathrm{auto}_{z,k}) / I(\theta_j; \mathrm{auto}_{z,k})$, which quantifies the relative importance of the cross-power measurement.
\begin{figure}[h!]
    \centering
    \includegraphics[width=0.9\linewidth]{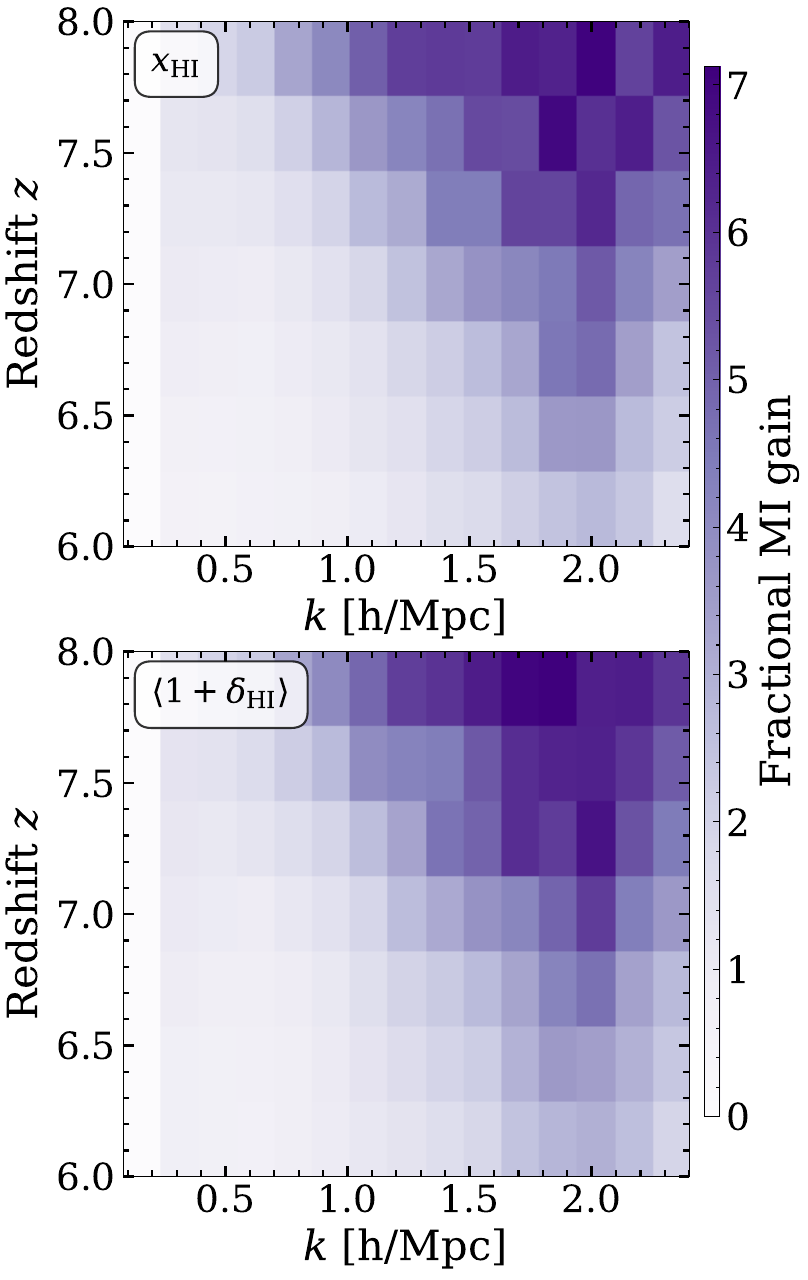}
    \caption{Fractional mutual information gain from including 21cm galaxy cross-power spectrum measurements in addition to 21cm auto-power ones, as a function of redshift and wave number. Upper panel: Neutral fraction, $x_\mathrm{HI}(z)$. Lower panel: Mean density contrast, $\langle 1+\delta_\mathrm{HI} \rangle(z)$. The color-scale indicates the ratio, $I(\theta; \mathrm{cross}_{z,k} | \mathrm{auto}_{z,k}) / I(\theta; \mathrm{auto}_{z,k})$, quantifying the relative additional information provided by cross-correlations. Purple regions indicate high information gain (cross-power provides substantial additional constraints), while white regions show minimal gain (auto-power is sufficient). The cross-power provides significant information across most scales and redshifts, with 78\% of bins showing fractional gains exceeding 100\%. The information gain increases with redshift and peaks on small scales $k \sim 1-2$ Mpc$^{-1}$, reflecting the evolving correlation between ionization morphology and galaxy clustering during reionization.}
    
    \label{fig:mi_fraction}
\end{figure}

We estimated these quantities directly from the simulation dataset. For the fiducial survey configuration (100 deg$^2$, $\sigma_z=0.001$, $ M_\mathrm{h,min}=10^{11}$ M$_\odot$), we took 1000 test simulations and extracted pairs of $(\theta,x)$, true parameter values, and their corresponding normalized power-spectrum features, using the normalization statistics stored in the trained flow model to ensure consistent preprocessing. 
For each $(z,k)$ bin, we estimated the marginal MI values $I(\theta_j; \mathrm{auto}_{z,k})$ and $I(\theta_j; \mathrm{cross}_{z,k})$, as well as the joint MI $I(\theta_j; [\mathrm{auto}_{z,k}, \mathrm{cross}_{z,k}])$, using a nonparametric discrete estimator. Both $\theta_j$ and each spectral feature were independently discretized into equiprobable quantile bins (32 bins for parameters and 15 bins for spectral features). The marginal MI was then computed from the empirical joint frequency table using Eq.~\eqref{eq:MI}. For the joint MI, the pair $(\mathrm{auto}_{z,k}, \mathrm{cross}_{z,k})$ was treated as a single compound discrete variable by encoding the two bin indices as a single index, making no assumption about independence between auto and cross features in the same $(z,k)$ bin. The conditional MI then follows from Eq.~\eqref{eq: CMI} by subtraction.
This estimator is entirely independent of the posterior approximation of the flow. The per-bin analysis treated different $(z,k)$ bins as being independent of one another; correlations between different bins were not captured. The conditional MI at a given bin therefore reflects the information of that bin marginally, not accounting for any redundancy or synergy with other bins.

Figure~\ref{fig:mi_fraction} shows the fractional MI gain for $x_\mathrm{HI}(z)$ (upper panel) and $\langle 1+\delta_\mathrm{HI} \rangle(z)$ (lower panel) as a function of redshift and scale. The color map indicates regions where the cross-power spectrum provides substantial additional information (purple) versus regions where the auto-power spectrum alone is sufficient (white). 

Several key trends emerge from this analysis.
First, the cross-power spectrum provides a significant conditional information gain across nearly all $(z,k)$ bins, with approximately 78\% of bins showing fractional gains exceeding 100\% for both parameters. The mean fractional gain is $\sim$270\% for $x_\mathrm{HI}$ and $\sim$280\% for $\langle 1+\delta_\mathrm{HI} \rangle$. This seemingly counter-intuitive result must be interpreted carefully. It does not imply that cross-power is more informative than auto-power in absolute terms. Section~\ref{sec: EoR timeline} demonstrated that 21cm auto-power alone provides the majority of constraints. Rather, these large fractional values reflect that cross-power measurements provide complementary information breaking specific degeneracies left unresolved by auto-power. 
While the marginal MI $I(\theta; \mathrm{cross})$ is relatively modest, the conditional MI $I(\theta; \mathrm{cross} | \mathrm{auto})$ can be substantial. Once the 21cm auto-power establishes the mean ionization state and overall neutral hydrogen distribution, the cross-correlation reveals the spatial relationship between ionization morphology and galaxy clustering; this information is particularly valuable for constraining astrophysical source properties and ionization morphology, which are only weakly determined by power-spectrum shape alone.

Second, there is a clear redshift dependence, with higher fractional gains toward earlier cosmic times (higher $z$). The fractional gain (averaged over k) increases monotonically from $\sim$135\% at $z=6.0$ to $\sim$450\% at $z=8.0$ for both parameters. This trend reflects the evolving character of reionization: at later times ($z \sim 6$), when reionization is well-advanced, the 21cm large-scale power and global ionization state provide strong constraints on the remaining parameters. At earlier times ($z \sim 8$) during the onset of reionization, the relationship between the first ionizing sources and the emerging ionized bubbles becomes increasingly critical, and the ability of cross-power to probe this connection yields proportionally larger information gains relative to what auto-power constrains.

Third, the scale dependence shows that large scales ($k \lesssim 0.1$ Mpc$^{-1}$) provide minimal information gains, with fractional MI values near zero. On these scales, cosmic variance dominates both the 21cm and galaxy signals, and their correlation adds little beyond what each observable independently constrains. The information gain increases systematically toward smaller scales, reaching $\sim$290\% (averaged over $z$) at $k \sim 1.2$ Mpc$^{-1}$ and $\sim$370\% on the smallest scales probed ($k \sim 2.4$ Mpc$^{-1}$). These intermediate to small scales are where ionization morphology, bubble sizes, shapes, and their correlation with the underlying density field traced by galaxies, are most directly manifested. The cross-power on these scales provides crucial constraints on how ionized regions grow around galaxies; this information complements the auto-power global characterization of the neutral fraction distribution.
Comparing the two parameters, $\langle 1+\delta_\mathrm{HI} \rangle$ shows slightly higher fractional gains than $x_\mathrm{HI}$ (by a factor of $\sim$1.03), though the difference is modest. The mean conditional MI values are 1.28 nats for $x_\mathrm{HI}$ and 1.30 nats for $\langle 1+\delta_\mathrm{HI} \rangle$, indicating comparable absolute information content from cross-power for both quantities.

These substantial conditional MI gains translate to modest but meaningful improvements in posterior precision for global reionization parameters, where auto-power already provides strong baseline constraints. The large conditional MI values on intermediate to small scales and at higher redshifts---precisely where ionization morphology and source-bubble correlations are most important---suggest that cross-power measurements carry crucial complementary information for breaking degeneracies unresolved by auto-power alone. This pattern is directly supported by gradient-based saliency analysis of the trained neural network (Appendix~\ref{app: saliency}), which reveals that the 21cm auto-power network focuses attention on low redshifts across all scales, while the cross-power network distributes attention across all redshifts but concentrates on intermediate scales. Since the 21cm auto-power already captures low-redshift information efficiently, the cross-power naturally provides the greatest information gain at higher redshifts where auto-power constraints are weaker, explaining why the conditional MI peaks precisely in these regimes. In the following section, we extend our analysis beyond the neutral fraction and density evolution to examine how well cross-power measurements constrain the astrophysical properties of the ionizing sources themselves.

\subsection{Inferring reionization source properties}\label{sec: astro}
To demonstrate the unique potential of 21cm galaxy cross-correlations for constraining source properties, we retrained our framework to infer the four astrophysical parameters, $\{f_{\mathrm{esc},10}, f_{*,10}, \alpha_\mathrm{esc}, \alpha_*\}$ (Eq.\eqref{eq: astro params}), from the power-spectrum measurements. Here $f_{\mathrm{esc},10}$ and $f_{*,10}$ denote the escape fraction and star formation efficiency normalized to halos with masses of $10^{10}$ M$_\odot$, while the power-law exponents describe their mass dependence. These parameters directly control where and how efficiently ionized bubbles form around galaxies, making them natural targets for cross-power inference but challenging to extract from 21cm auto-power alone, which primarily traces the integrated ionization state rather than source-bubble correlations.
\begin{figure*}[h!]
\sidecaption
    \centering
    \includegraphics[width=12cm]{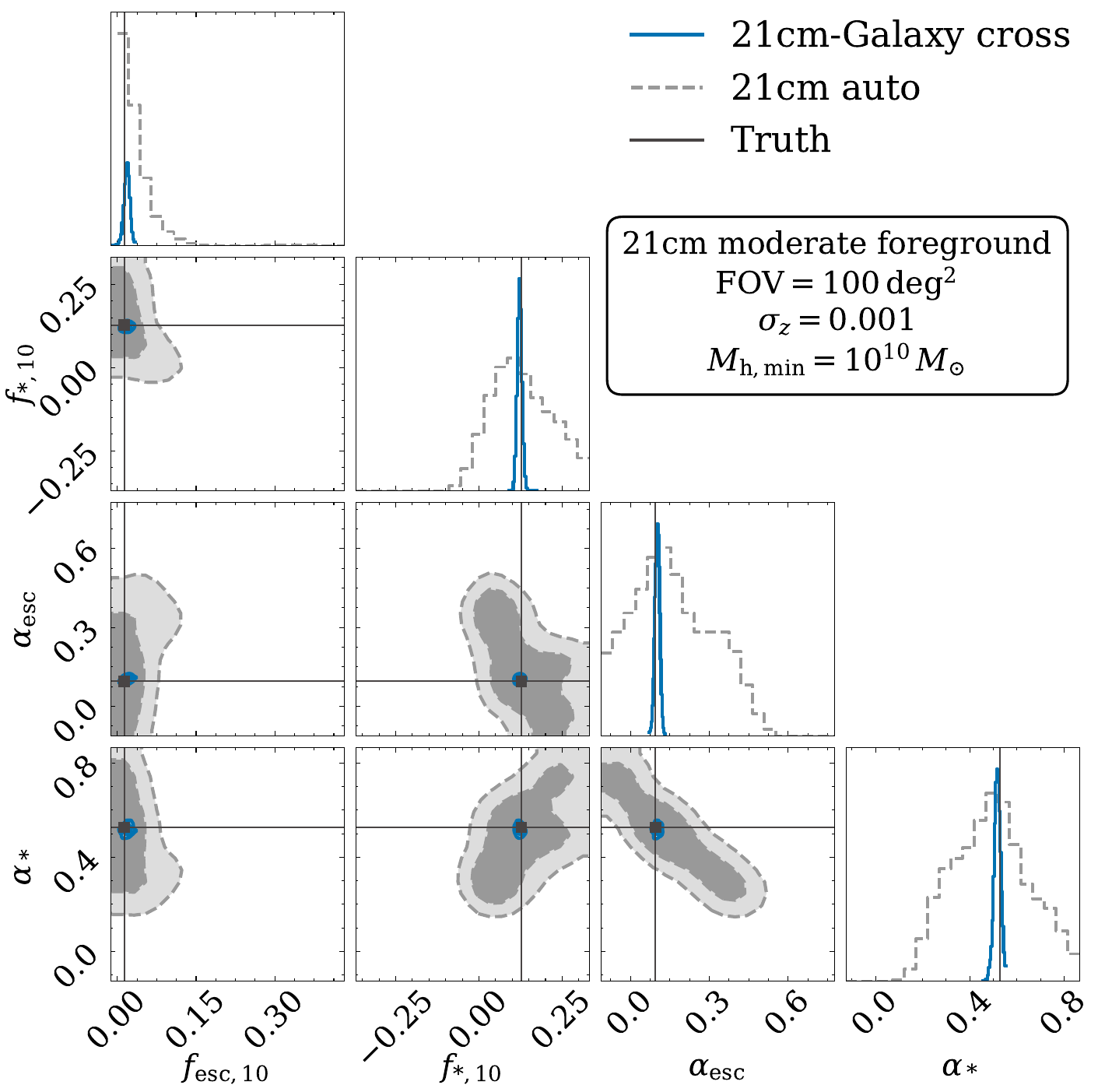}
    \caption{Inference of astrophysical source parameters from 21cm auto-power and 21cm galaxy cross-power spectra for a randomly chosen sample assuming moderate 21cm foreground avoidance. Corner plots show the marginalized posteriors for the four parameters controlling escape fraction and star formation efficiency: $\log_{10} f_{\mathrm{esc},10}$, $\log_{10} f_{*,10}$, $\alpha_\mathrm{esc}$, and $\alpha_*$. Solid blue contours show cross-power constraints (R$^2 > 0.92$, mean posterior volume of $\sim$11\% across the entire test dataset). Dashed gray contours show auto-power constraints, which are virtually uninformative (R$^2 < 0.47$, posterior volume of $>60\%$ across the entire test dataset). Results assume an optimistic future survey with a FOV of  100 deg$^2$, spectroscopic redshift uncertainty of $\sigma_z=0.001,$ and a halo mass detection threshold of $M_\mathrm{h,min} = 10^{10}$ M$_\odot$ (for comparison with optimistic foreground removal at $M_\mathrm{h,min} = 10^{11}$ M$_\odot$, see Fig.~\ref{fig:astro_corner_opt}). The dramatic contrast demonstrates that cross-correlations provide access to source properties fundamentally inaccessible to 21cm auto-power alone, though this requires detecting faint galaxies below currently planned survey sensitivities.}
    \label{fig:astro_corner}
\end{figure*}

Fig.~\ref{fig:astro_corner} presents the inference results as an example corner plot using the fiducial survey configuration (100 deg$^2$, $\sigma_z=0.001$) but with an optimistic halo mass threshold of $ M_\mathrm{h,min} = 10^{10}$ M$_\odot$. This choice probes the deepest future galaxy surveys and serves as a proof of concept for the method's potential. The 21cm auto-power spectrum (dashed gray contours) shows no constraining power beyond revealing strong parameter degeneracies, particularly between $\alpha_*$ and $\alpha_\mathrm{esc}$ where different ionization morphologies yield identical EoR timelines. Across the whole test set the R$^2$ scores (where 1 indicates perfect inference accuracy) range from 0.08 to 0.47, and the relative PV is between 60\% and 87\%. The posteriors remain broad and weakly correlated with the true values, indicating that the auto-power spectrum alone, while highly informative of the neutral fraction evolution, carries insufficient information to break degeneracies among source parameters that produce similar global ionization histories through different combinations of escape fractions and star formation efficiencies.

In stark contrast, the 21cm galaxy cross-power spectrum (solid blue contours) provides excellent constraints on all four parameters. Tight, circular contours indicate precise and nondegenerate constraints, accurately recovering the true parameters. Across the test dataset the R$^2$ scores consistently exceed 0.92, and the mean relative PV is $\sim$11\% across all quantities. This dramatic improvement can be understood physically. While the auto-power measures the spatial distribution of neutral hydrogen, the cross-power directly probes the correlation between ionized regions and galaxy positions. This correlation is highly sensitive to how ionizing photons escape from galaxies (through $f_\mathrm{esc}$) and how efficiently stars form in halos of different masses (through $f_*$). Galaxies with higher escape fractions create larger ionized bubbles around themselves, strengthening the anti-correlation signal in the cross-power, while the mass dependencies determine which galaxy populations contribute most to reionization and thus shape the scale dependence of the cross-power.

Combining both observables (cross+auto) yields performance levels comparable to that of cross-power alone on average, though with substantially larger run-to-run variance ($\mathrm{PV}=0.24\pm0.12$ compared to $0.11\pm0.02$ for cross-power alone). This increased epistemic noise likely reflects the challenge of training a single neural density estimator on two observables with vastly different information; the cross-power is highly informative of source 
properties, while the auto-power is not, and the network must allocate its capacity to model the joint likelihood of both. The large variance suggests that the network sometimes successfully focuses on the informative cross-power features and sometimes does not, depending on initialization. A dedicated architecture optimized for the combined input, or an adaptive weighting scheme, would likely yield more consistent results. We quantified the epistemic 
uncertainty by retraining each model across multiple random initializations. 
The 21cm auto-power models show negligible run-to-run variance 
($\mathrm{PV}=0.76\pm0.01$), consistent with all initializations converging to the same uninformative posterior. The cross-power models also show small variance ($\mathrm{PV}=0.11\pm0.02$), with all qualitative conclusions remaining robust across runs. The epistemic uncertainty across all configurations is discussed further in Appendix~\ref{app: stochasticity}.

The results presented here assume an optimistic $M_\mathrm{h,min} = 10^{10}$ M$_\odot$ halo mass threshold, corresponding to extremely deep future galaxy surveys that can detect faint star-forming galaxies down to this limit across 100 deg$^2$. Increasing the threshold to $M_\mathrm{h,min} = 10^{10.5}$ M$_\odot$ yields only moderately degraded constraints ($\mathrm{PV}=0.17\pm0.03$), suggesting that surveys reaching this intermediate depth already provide meaningful source-property constraints. At the more commonly assumed $M_\mathrm{h,min} = 10^{11}$ M$_\odot$ threshold however, the source parameter constraints become marginal ($\mathrm{PV}=0.50\pm0.10$). On the contrary, Appendix~\ref{app: optimistic} demonstrates that optimistic 21cm foreground removal strategies, which recover wedge modes, can restore tight constraints on source properties ($\mathrm{PV}=0.18\pm 0.04$) even at the $M_\mathrm{h,min} = 10^{11}$ M$_\odot$ threshold, offering a complementary pathway to achieving source-property constraints without requiring detection of the faintest galaxies. We note that additional uncertainties arise when detecting galaxies via, for example, Lyman-$\alpha$ emission, including the stochasticity of luminosities at fixed halo mass and variations in IGM transmission. These effects contribute to the effective shot noise and scatter in galaxy bias. However, for typical survey sensitivities, such contributions remain subdominant compared to uncertainties from survey completeness, set by the halo mass (or luminosity) detection threshold, and 21cm signal foreground contamination.

The contrast between Fig.~\ref{fig:mi_fraction} (showing large conditional MI for neutral fraction and density) and Fig.~\ref{fig:astro_corner} (showing weak auto-power constraints on source properties) illustrates a key distinction. While cross-power provides complementary information that modestly improves constraints on parameters already well measured by auto-power, it provides essential information for parameters that are degenerate or invisible to auto-power alone. This dual role of enhancement and enablement positions 21cm galaxy cross-correlations as a uniquely powerful probe of cosmic reionization. By simultaneously confirming the cosmological 21cm signal, providing robustness against foreground contamination, and accessing source properties inaccessible to 21cm measurements alone, cross-power studies represent not merely an incremental improvement over single-probe analyses, but a qualitatively new observational window into the EoR. 

\section{Conclusions}\label{sec: conclusion}
In this work, we developed an SBI framework to quantify the constraining power of 21cm galaxy cross-correlations during the EoR.
Our methodology combines realistic forward modeling of noisy, 21cm galaxy cross-power spectra generated with \texttt{21cmFASTv4} and incorporates instrumental and survey effects following~\citet{Hutter_2025}, with the \texttt{EoRFlow} SBI framework of~\citet{Pietschke_2025}. This approach enables likelihood-free inference on astrophysical parameters governing reionization without requiring analytic approximations to the intractable data likelihood. The inference network operates directly on spherically averaged 1D power spectra using conditional normalizing flows with rational quadratic spline coupling layers, providing increased expressivity compared to earlier implementations. Our main findings are as follows.

\paragraph{Inference of global reionization properties:}Assuming 100h of SKA AA* observations with moderate foreground avoidance and a WST-type fiducial galaxy survey ($\mathrm{FOV}=100\mathrm{deg}^2$, $\sigma_z=0.001$, $M_\mathrm{h,min}=10^{11}\mathrm{M}_\odot$), we demonstrate that 21cm galaxy cross-power spectra alone yield unbiased constraints on the neutral hydrogen fraction $x_\mathrm{HI}(z)$ and the mean overdensity in neutral regions $\langle 1+\delta_\mathrm{HI} \rangle(z)$ across $z=6$-$8$. While these constraints are weaker than those obtained from the 21cm auto-power spectrum, they provide a physically independent probe of the reionization history. When combined with auto-power measurements, posterior volumes decrease by approximately 20--30\% compared to auto-power ones alone, confirming that cross-correlations contribute genuinely complementary information rather than redundant constraints.
\paragraph{Galaxy-survey requirements:} Our systematic exploration of survey parameter space reveals that redshift precision is the dominant factor determining the scientific utility of cross-correlation measurements. Spectroscopic-quality redshifts ($\sigma_z \sim 0.001$) provide optimal constraints. Intermediate precision achievable with grisms ($\sigma_z \sim 0.01$) can yield 
meaningful constraints when combined with large survey areas, though these configurations carry the largest epistemic uncertainty and their PV values should be interpreted qualitatively. In all cases they fall short of the factor of $\sim$10 improvement achievable with full spectroscopy. Photometric uncertainties ($\sigma_z \sim 0.1$) severely degrade constraints regardless of survey area or depth. FOV and minimum detectable halo mass play secondary but important roles, with deeper surveys ($M_\mathrm{h,min} \lesssim 10^{10.5}\mathrm{M}_\odot$) partially compensating for smaller survey areas. These findings provide concrete guidance for designing future galaxy surveys intended for cross-correlation studies with 21cm experiments.
\paragraph{Complementary information from cross-correlations:} Using mutual information analysis, we quantified where cross-power measurements provide the greatest information gain beyond what the 21cm auto-power spectrum constrains. The additional information is concentrated at intermediate to small scales ($k \sim 0.5$-$2\mathrm{Mpc}^{-1}$), corresponding to characteristic ionized bubble sizes, and increases toward higher redshifts during the early phases of reionization. This pattern reflects the sensitivity of cross-correlations to the evolving spatial relationship between ionizing sources and the surrounding ionization morphology.
\paragraph{Constraints on ionizing source properties:} The most significant finding is that 21cm galaxy cross-correlations provide access to astrophysical source properties that are fundamentally inaccessible to 21cm auto-power measurements alone. We showed that the escape fraction ($f_{\mathrm{esc},10}$, $\alpha_\mathrm{esc}$) as well as the star formation efficiency parameters ($f_{*,10}$, $\alpha_*$) can be tightly constrained from cross-power spectra, achieving R$^2 > 0.92$ and mean posterior volumes of $\sim$11\%. In contrast, the 21cm auto-power spectrum yields R$^2 < 0.5$ and posterior volumes exceeding 60\%, owing to strong degeneracies among source models that produce similar global ionization histories. Notably, this unique constraining power of 21cm galaxy cross-correlations can be achieved through either deep surveys detecting faint galaxies down to $M_\mathrm{h,min} \sim 10^{10}$ M$_\odot$ under moderate foreground treatment, or more conservative mass thresholds ($M_\mathrm{h,min} \sim 10^{11}$ M$_\odot$) when paired with optimistic 21cm foreground removal (Appendix~\ref{app: optimistic}). This capability transforms cross-correlations into a powerful diagnostic of the reionization source population.
\paragraph{Implications for observational programs:} Our results carry direct implications for the planning and prioritization of forthcoming surveys. For facilities such as the WST and the PFS, spectroscopic observations over areas of $\gtrsim 10$-$100\mathrm{deg}^2$ with high redshift precision will be essential to exploit the full potential of 21cm galaxy cross-correlations. Critically, tight constraints on astrophysical source properties can be achieved through either deep surveys reaching $M_\mathrm{h,min} \sim 10^{10}$-$10^{10.5}\mathrm{M}_\odot$ under moderate 21cm foreground treatment, or surveys targeting brighter galaxies ($M_\mathrm{h,min} \sim 10^{11}\mathrm{M}_\odot$) paired with optimistic foreground removal strategies (Appendix~\ref{app: optimistic}). This flexibility allows observational programs to optimize either galaxy survey sensitivity or 21cm foreground cleaning capabilities depending on technological developments and resource allocation.
\paragraph{Future directions:} This work establishes a foundation for inference on 21cm galaxy cross-correlations, but several extensions promise further scientific gains. Our analysis assumed moderate 21cm foreground avoidance. Optimistic foreground removal strategies could substantially enhance constraining power. As demonstrated in Appendix~\ref{app: optimistic}, this relaxation enables tight constraints on astrophysical source properties even when galaxy surveys reach only $M_\mathrm{h,min} \sim 10^{11}$ M$_\odot$. These findings align with the analysis of~\citet{Hutter_2025}, which showed that optimistic foreground scenarios dramatically relax survey requirements for detecting the cross-correlation signal itself. The 1D power spectrum employed here, while providing a favorable S/N, discards information encoded in the anisotropic and non-Gaussian structure of the cross-correlation signal. Moving beyond power spectra to leverage topological summary statistics, such as Minkowski functionals or Betti numbers, could capture the morphological information encoded in the shapes and connectivity of ionized regions. Machine-learning approaches offer a natural pathway to extract such features, either through learned summary networks that compress the full 3D cross-correlation field or through neural-network architectures designed to operate directly on spatial data. 

In summary, 21cm galaxy cross-correlations occupy a unique position in the landscape of reionization probes. They provide an independent confirmation of the cosmological 21cm signal, are robust against foreground contamination, and unlock constraints on source properties that are inaccessible to 21cm auto-power measurements alone. As both 21cm experiments and high-redshift galaxy surveys approach the sensitivities required for cross-correlation detections, the SBI framework developed here offers a flexible and scalable approach to extracting maximal astrophysical insight from this powerful multi-tracer view of the EoR.

\section*{Data availability}
The code used to compute the 21cm galaxy cross-power spectrum uncertainties is publicly available at \href{https://github.com/annehutter/21cm_gal_uncertainties}{github.com/annehutter/21cm\_gal\_uncertainties}. This work builds upon the \texttt{EoRFlow} framework \citep{Pietschke_2025}, available at \href{https://github.com/astro-ML/EoRFlow}{github.com/astro-ML/EoRFlow}. The project-specific analysis scripts and trained neural-network models can be found at \href{https://github.com/astro-ML/21cmgalaxy-SBI}{github.com/astro-ML/21cmgalaxy-SBI}. 

\begin{acknowledgements}
We would like to thank Maike Voelkel, Vrund Patel and M. Paola Vaccaro for helpful discussions. We also thank the anonymous referee for their insightful comments, which improved the clarity and robustness of this work.
YP's and CH's work is funded by the Volkswagen Foundation. This work was supported by the DFG under Germany's Excellence Strategy EXC 2181/1 - 390900948 The Heidelberg STRUCTURES Excellence Cluster. The authors acknowledge support by the High Performance and Cloud Computing Group at the Zentrum für Datenverarbeitung of the University of Tübingen, the state of Baden-Württemberg through bwHPC and the German Research Foundation (DFG) through grant no INST 37/935-1 FUGG. 
\end{acknowledgements}

\bibliographystyle{aa} 
\bibliography{references} 

@ARTICLE{Furlanetto2007,
       author = {{Furlanetto}, Steven R. and {Lidz}, Adam},
        title = "{The Cross-Correlation of High-Redshift 21 cm and Galaxy Surveys}",
      journal = {\apj},
         year = 2007,
        month = may,
       volume = {660},
       number = {2},
        pages = {1030-1038},
          doi = {10.1086/513009},
archivePrefix = {arXiv},
       eprint = {astro-ph/0611274},
 primaryClass = {astro-ph},
       adsurl = {https://ui.adsabs.harvard.edu/abs/2007ApJ...660.1030F}
}

@ARTICLE{Furlanetto2006,
       author = {{Furlanetto}, Steven R. and {Oh}, S. Peng and {Briggs}, Frank H.},
        title = "{Cosmology at low frequencies: The 21 cm transition and the high-redshift Universe}",
      journal = {\physrep},
         year = 2006,
        month = oct,
       volume = {433},
       number = {4-6},
        pages = {181-301},
          doi = {10.1016/j.physrep.2006.08.002},
archivePrefix = {arXiv},
       eprint = {astro-ph/0608032},
 primaryClass = {astro-ph},
       adsurl = {https://ui.adsabs.harvard.edu/abs/2006PhR...433..181F}
}

@book{Cover2006,
  title={Elements of Information Theory},
  author={Cover, Thomas M. and Thomas, Joy A.},
  year={2006},
  edition={2nd},
  publisher={Wiley-Interscience},
  address={Hoboken, NJ}
}

@ARTICLE{Schosser_2026,
       author = {{Schosser}, Benedikt and {Heneka}, Caroline and {Sch{\"a}fer}, Bj{\"o}rn Malte},
        title = "{Starobinsky in stereo: SKA-CMB synergy in SBI}",
      journal = {JCAP},
         year = 2026,
        month = feb,
       volume = {2026},
       number = {2},
          eid = {086},
        pages = {086},
          doi = {10.1088/1475-7516/2026/02/086},
archivePrefix = {arXiv},
       eprint = {2508.10094},
 primaryClass = {astro-ph.CO},
       adsurl = {https://ui.adsabs.harvard.edu/abs/2026JCAP...02..086S}
}

@ARTICLE{Gagnon-Hartman2025,
       author = {{Gagnon-Hartman}, Samuel and {Davies}, James E. and {Mesinger}, Andrei},
        title = "{Detecting galaxy {\textendash} 21-cm cross-correlation during reionization}",
      journal = {\aap},
         year = 2025,
        month = jul,
       volume = {699},
          eid = {A131},
        pages = {A131},
          doi = {10.1051/0004-6361/202554456},
archivePrefix = {arXiv},
       eprint = {2502.20447},
 primaryClass = {astro-ph.CO},
       adsurl = {https://ui.adsabs.harvard.edu/abs/2025A&A...699A.131G}
}

@ARTICLE{Heneka2017,
       author = {{Heneka}, Caroline and {Cooray}, Asantha and {Feng}, Chang},
        title = "{Probing the Intergalactic Medium with Ly{\ensuremath{\alpha}} and 21 cm Fluctuations}",
      journal = {\apj},
         year = 2017,
        month = oct,
       volume = {848},
       number = {1},
          eid = {52},
        pages = {52},
          doi = {10.3847/1538-4357/aa8eed},
archivePrefix = {arXiv},
       eprint = {1611.09682},
 primaryClass = {astro-ph.CO},
       adsurl = {https://ui.adsabs.harvard.edu/abs/2017ApJ...848...52H}
}

@ARTICLE{Heneka2020,
       author = {{Heneka}, Caroline and {Mesinger}, Andrei},
        title = "{The spin-temperature dependence of the 21-cm-LAE cross-correlation}",
      journal = {\mnras},
         year = 2020,
        month = jul,
       volume = {496},
       number = {1},
        pages = {581-589},
          doi = {10.1093/mnras/staa1517},
archivePrefix = {arXiv},
       eprint = {2004.10097},
 primaryClass = {astro-ph.CO},
       adsurl = {https://ui.adsabs.harvard.edu/abs/2020MNRAS.496..581H}
}

@ARTICLE{Heneka2021,
       author = {{Heneka}, Caroline and {Cooray}, Asantha},
        title = "{Optimal survey parameters: Ly {\ensuremath{\alpha}} and H {\ensuremath{\alpha}} intensity mapping for synergy with the 21-cm signal during reionization}",
      journal = {\mnras},
         year = 2021,
        month = sep,
       volume = {506},
       number = {2},
        pages = {1573-1584},
          doi = {10.1093/mnras/stab1842},
archivePrefix = {arXiv},
       eprint = {2104.12739},
 primaryClass = {astro-ph.CO},
       adsurl = {https://ui.adsabs.harvard.edu/abs/2021MNRAS.506.1573H}
}

@ARTICLE{Kubota2018,
       author = {{Kubota}, Kenji and {Yoshiura}, Shintaro and {Takahashi}, Keitaro and {Hasegawa}, Kenji and {Yajima}, Hidenobu and {Ouchi}, Masami and {Pindor}, B. and {Webster}, R.~L.},
        title = "{Detectability of the 21-cm signal during the epoch of reionization with 21-cm Lyman {\ensuremath{\alpha}} emitter cross-correlation - I}",
      journal = {\mnras},
         year = 2018,
        month = sep,
       volume = {479},
       number = {2},
        pages = {2754-2766},
          doi = {10.1093/mnras/sty1471},
archivePrefix = {arXiv},
       eprint = {1708.06291},
 primaryClass = {astro-ph.CO},
       adsurl = {https://ui.adsabs.harvard.edu/abs/2018MNRAS.479.2754K}
}

@ARTICLE{HERA2022,
       author = {{Abdurashidova}, Zara and {Aguirre}, James E. and {Alexander}, Paul and {Ali}, Zaki S. and {Balfour}, Yanga and {Barkana}, Rennan and {Beardsley}, Adam P. and {Bernardi}, Gianni and {Billings}, Tashalee S. and {Bowman}, Judd D. and {Bradley}, Richard F. and {Bull}, Philip and {Burba}, Jacob and {Carey}, Steve and {Carilli}, Chris L. and {Cheng}, Carina and {DeBoer}, David R. and {Dexter}, Matt and {de Lera Acedo}, Eloy and {Dillon}, Joshua S. and {Ely}, John and {Ewall-Wice}, Aaron and {Fagnoni}, Nicolas and {Fialkov}, Anastasia and {Fritz}, Randall and {Furlanetto}, Steven R. and {Gale-Sides}, Kingsley and {Glendenning}, Brian and {Gorthi}, Deepthi and {Greig}, Bradley and {Grobbelaar}, Jasper and {Halday}, Ziyaad and {Hazelton}, Bryna J. and {Heimersheim}, Stefan and {Hewitt}, Jacqueline N. and {Hickish}, Jack and {Jacobs}, Daniel C. and {Julius}, Austin and {Kern}, Nicholas S. and {Kerrigan}, Joshua and {Kittiwisit}, Piyanat and {Kohn}, Saul A. and {Kolopanis}, Matthew and {Lanman}, Adam and {La Plante}, Paul and {Lekalake}, Telalo and {Lewis}, David and {Liu}, Adrian and {Ma}, Yin-Zhe and {MacMahon}, David and {Malan}, Lourence and {Malgas}, Cresshim and {Maree}, Matthys and {Martinot}, Zachary E. and {Matsetela}, Eunice and {Mesinger}, Andrei and {Mirocha}, Jordan and {Molewa}, Mathakane and {Morales}, Miguel F. and {Mosiane}, Tshegofalang and {Mu{\~n}oz}, Julian B. and {Murray}, Steven G. and {Neben}, Abraham R. and {Nikolic}, Bojan and {Nunhokee}, Chuneeta D. and {Parsons}, Aaron R. and {Patra}, Nipanjana and {Pieterse}, Samantha and {Pober}, Jonathan C. and {Qin}, Yuxiang and {Razavi-Ghods}, Nima and {Reis}, Itamar and {Ringuette}, Jon and {Robnett}, James and {Rosie}, Kathryn and {Santos}, Mario G. and {Sikder}, Sudipta and {Sims}, Peter and {Smith}, Craig and {Syce}, Angelo and {Thyagarajan}, Nithyanandan and {Williams}, Peter K.~G. and {Zheng}, Haoxuan},
        title = "{HERA Phase I Limits on the Cosmic 21 cm Signal: Constraints on Astrophysics and Cosmology during the Epoch of Reionization}",
      journal = {\apj},
         year = 2022,
        month = jan,
       volume = {924},
       number = {2},
          eid = {51},
        pages = {51},
          doi = {10.3847/1538-4357/ac2ffc},
archivePrefix = {arXiv},
       eprint = {2108.07282},
 primaryClass = {astro-ph.CO},
       adsurl = {https://ui.adsabs.harvard.edu/abs/2022ApJ...924...51A}
}

@ARTICLE{HERA2023,
       author = {{HERA Collaboration} and {Abdurashidova}, Zara and {Adams}, Tyrone and {Aguirre}, James E. and {Alexander}, Paul and {Ali}, Zaki S. and {Baartman}, Rushelle and {Balfour}, Yanga and {Barkana}, Rennan and {Beardsley}, Adam P. and {Bernardi}, Gianni and {Billings}, Tashalee S. and {Bowman}, Judd D. and {Bradley}, Richard F. and {Breitman}, Daniela and {Bull}, Philip and {Burba}, Jacob and {Carey}, Steve and {Carilli}, Chris L. and {Cheng}, Carina and {Choudhuri}, Samir and {DeBoer}, David R. and {de Lera Acedo}, Eloy and {Dexter}, Matt and {Dillon}, Joshua S. and {Ely}, John and {Ewall-Wice}, Aaron and {Fagnoni}, Nicolas and {Fialkov}, Anastasia and {Fritz}, Randall and {Furlanetto}, Steven R. and {Gale-Sides}, Kingsley and {Garsden}, Hugh and {Glendenning}, Brian and {Gorce}, Ad{\'e}lie and {Gorthi}, Deepthi and {Greig}, Bradley and {Grobbelaar}, Jasper and {Halday}, Ziyaad and {Hazelton}, Bryna J. and {Heimersheim}, Stefan and {Hewitt}, Jacqueline N. and {Hickish}, Jack and {Jacobs}, Daniel C. and {Julius}, Austin and {Kern}, Nicholas S. and {Kerrigan}, Joshua and {Kittiwisit}, Piyanat and {Kohn}, Saul A. and {Kolopanis}, Matthew and {Lanman}, Adam and {La Plante}, Paul and {Lewis}, David and {Liu}, Adrian and {Loots}, Anita and {Ma}, Yin-Zhe and {MacMahon}, David H.~E. and {Malan}, Lourence and {Malgas}, Keith and {Malgas}, Cresshim and {Maree}, Matthys and {Marero}, Bradley and {Martinot}, Zachary E. and {McBride}, Lisa and {Mesinger}, Andrei and {Mirocha}, Jordan and {Molewa}, Mathakane and {Morales}, Miguel F. and {Mosiane}, Tshegofalang and {Mu{\~n}oz}, Julian B. and {Murray}, Steven G. and {Nagpal}, Vighnesh and {Neben}, Abraham R. and {Nikolic}, Bojan and {Nunhokee}, Chuneeta D. and {Nuwegeld}, Hans and {Parsons}, Aaron R. and {Pascua}, Robert and {Patra}, Nipanjana and {Pieterse}, Samantha and {Qin}, Yuxiang and {Razavi-Ghods}, Nima and {Robnett}, James and {Rosie}, Kathryn and {Santos}, Mario G. and {Sims}, Peter and {Singh}, Saurabh and {Smith}, Craig and {Swarts}, Hilton and {Tan}, Jianrong and {Thyagarajan}, Nithyanandan and {Wilensky}, Michael J. and {Williams}, Peter K.~G. and {van Wyngaarden}, Pieter and {Zheng}, Haoxuan},
        title = "{Improved Constraints on the 21 cm EoR Power Spectrum and the X-Ray Heating of the IGM with HERA Phase I Observations}",
      journal = {\apj},
         year = 2023,
        month = mar,
       volume = {945},
       number = {2},
          eid = {124},
        pages = {124},
          doi = {10.3847/1538-4357/acaf50},
archivePrefix = {arXiv},
       eprint = {2210.04912},
 primaryClass = {astro-ph.CO},
       adsurl = {https://ui.adsabs.harvard.edu/abs/2023ApJ...945..124H}
}

@ARTICLE{LaPlante2023,
       author = {{La Plante}, Paul and {Mirocha}, Jordan and {Gorce}, Ad{\'e}lie and {Lidz}, Adam and {Parsons}, Aaron},
        title = "{Prospects for 21 cm Galaxy Cross-correlations with HERA and the Roman High-latitude Survey}",
      journal = {\apj},
         year = 2023,
        month = feb,
       volume = {944},
       number = {1},
          eid = {59},
        pages = {59},
          doi = {10.3847/1538-4357/acaeb0},
archivePrefix = {arXiv},
       eprint = {2205.09770},
 primaryClass = {astro-ph.CO},
       adsurl = {https://ui.adsabs.harvard.edu/abs/2023ApJ...944...59L}
}

@ARTICLE{Lidz2009,
       author = {{Lidz}, Adam and {Zahn}, Oliver and {Furlanetto}, Steven R. and {McQuinn}, Matthew and {Hernquist}, Lars and {Zaldarriaga}, Matias},
        title = "{Probing Reionization with the 21 cm Galaxy Cross-Power Spectrum}",
      journal = {\apj},
         year = 2009,
        month = jan,
       volume = {690},
       number = {1},
        pages = {252-266},
          doi = {10.1088/0004-637X/690/1/252},
archivePrefix = {arXiv},
       eprint = {0806.1055},
 primaryClass = {astro-ph},
       adsurl = {https://ui.adsabs.harvard.edu/abs/2009ApJ...690..252L}
}

@ARTICLE{Park2014,
       author = {{Park}, Jaehong and {Kim}, Han-Seek and {Wyithe}, J. Stuart B. and {Lacey}, C.~G.},
        title = "{The cross-power spectrum between 21 cm emission and galaxies in hierarchical galaxy formation models}",
      journal = {\mnras},
         year = 2014,
        month = mar,
       volume = {438},
       number = {3},
        pages = {2474-2482},
          doi = {10.1093/mnras/stt2366},
archivePrefix = {arXiv},
       eprint = {1309.3350},
 primaryClass = {astro-ph.CO},
       adsurl = {https://ui.adsabs.harvard.edu/abs/2014MNRAS.438.2474P}
}

@ARTICLE{Vrbanec2016,
       author = {{Vrbanec}, Dijana and {Ciardi}, Benedetta and {Jeli{\'c}}, Vibor and {Jensen}, Hannes and {Zaroubi}, Saleem and {Fernandez}, Elizabeth R. and {Ghosh}, Abhik and {Iliev}, Ilian T. and {Kakiichi}, Koki and {Koopmans}, L{\'e}on V.~E. and {Mellema}, Garrelt},
        title = "{Predictions for the 21 cm-galaxy cross-power spectrum observable with LOFAR and Subaru}",
      journal = {\mnras},
         year = 2016,
        month = mar,
       volume = {457},
       number = {1},
        pages = {666-675},
          doi = {10.1093/mnras/stv2993},
archivePrefix = {arXiv},
       eprint = {1509.03464},
 primaryClass = {astro-ph.CO},
       adsurl = {https://ui.adsabs.harvard.edu/abs/2016MNRAS.457..666V}
}

@ARTICLE{Vrbanec2020,
       author = {{Vrbanec}, Dijana and {Ciardi}, Benedetta and {Jeli{\'c}}, Vibor and {Jensen}, Hannes and {Iliev}, Ilian T. and {Mellema}, Garrelt and {Zaroubi}, Saleem},
        title = "{Predictions for the 21cm-galaxy cross-power spectrum observable with SKA and future galaxy surveys}",
      journal = {\mnras},
         year = 2020,
        month = mar,
       volume = {492},
       number = {4},
        pages = {4952-4958},
          doi = {10.1093/mnras/staa183},
archivePrefix = {arXiv},
       eprint = {2001.08814},
 primaryClass = {astro-ph.CO},
       adsurl = {https://ui.adsabs.harvard.edu/abs/2020MNRAS.492.4952V}
}

@ARTICLE{Weinberger2020,
       author = {{Weinberger}, Lewis H. and {Kulkarni}, Girish and {Haehnelt}, Martin G.},
        title = "{Probing delayed-end reionization histories with the 21-cm LAE cross-power spectrum}",
      journal = {\mnras},
         year = 2020,
        month = may,
       volume = {494},
       number = {1},
        pages = {703-718},
          doi = {10.1093/mnras/staa749},
archivePrefix = {arXiv},
       eprint = {1911.11783},
 primaryClass = {astro-ph.CO},
       adsurl = {https://ui.adsabs.harvard.edu/abs/2020MNRAS.494..703W}
}

@article{Planck2018,
    author = "Aghanim, N. and others",
    collaboration = "Planck",
    title = "{Planck 2018 results. VI. Cosmological parameters}",
    eprint = "1807.06209",
    archivePrefix = "arXiv",
    primaryClass = "astro-ph.CO",
    doi = "10.1051/0004-6361/201833910",
    journal = "\aap",
    volume = "641",
    pages = "A6",
    year = "2020",
    note = "[Erratum: A\&A 652, C4 (2021)]"
}

@article{Parsons_2010,
doi = {10.1088/0004-6256/139/4/1468},
url = {https://dx.doi.org/10.1088/0004-6256/139/4/1468},
year = {2010},
month = {mar},
publisher = {The American Astronomical Society},
volume = {139},
number = {4},
pages = {1468},
author = {Aaron R. Parsons and Donald C. Backer and Griffin S. Foster and Melvyn C. H. Wright and Richard F. Bradley and Nicole E. Gugliucci and Chaitali R. Parashare and Erin E. Benoit and James E. Aguirre and Daniel C. Jacobs and Chris L. Carilli and David Herne and Mervyn J. Lynch and Jason R. Manley and Daniel J. Werthimer},
title = {THE PRECISION ARRAY FOR PROBING THE EPOCH OF RE-IONIZATION: EIGHT STATION RESULTS},
journal = {\aj}
}

@ARTICLE{Tingay_2013,
       author = {{Tingay}, S.~J. and {Goeke}, R. and {Bowman}, J.~D. and {Emrich}, D. and {Ord}, S.~M. and {Mitchell}, D.~A. and {Morales}, M.~F. and {Booler}, T. and {Crosse}, B. and {Wayth}, R.~B. and {Lonsdale}, C.~J. and {Tremblay}, S. and {Pallot}, D. and {Colegate}, T. and {Wicenec}, A. and {Kudryavtseva}, N. and {Arcus}, W. and {Barnes}, D. and {Bernardi}, G. and {Briggs}, F. and {Burns}, S. and {Bunton}, J.~D. and {Cappallo}, R.~J. and {Corey}, B.~E. and {Deshpande}, A. and {Desouza}, L. and {Gaensler}, B.~M. and {Greenhill}, L.~J. and {Hall}, P.~J. and {Hazelton}, B.~J. and {Herne}, D. and {Hewitt}, J.~N. and {Johnston-Hollitt}, M. and {Kaplan}, D.~L. and {Kasper}, J.~C. and {Kincaid}, B.~B. and {Koenig}, R. and {Kratzenberg}, E. and {Lynch}, M.~J. and {Mckinley}, B. and {Mcwhirter}, S.~R. and {Morgan}, E. and {Oberoi}, D. and {Pathikulangara}, J. and {Prabu}, T. and {Remillard}, R.~A. and {Rogers}, A.~E.~E. and {Roshi}, A. and {Salah}, J.~E. and {Sault}, R.~J. and {Udaya-Shankar}, N. and {Schlagenhaufer}, F. and {Srivani}, K.~S. and {Stevens}, J. and {Subrahmanyan}, R. and {Waterson}, M. and {Webster}, R.~L. and {Whitney}, A.~R. and {Williams}, A. and {Williams}, C.~L. and {Wyithe}, J.~S.~B.},
        title = "{The Murchison Widefield Array: The Square Kilometre Array Precursor at Low Radio Frequencies}",
      journal = {\pasa},
         year = 2013,
        month = jan,
       volume = {30},
          eid = {e007},
        pages = {e007},
          doi = {10.1017/pasa.2012.007},
archivePrefix = {arXiv},
       eprint = {1206.6945},
 primaryClass = {astro-ph.IM},
       adsurl = {https://ui.adsabs.harvard.edu/abs/2013PASA...30....7T}
}

@article{LOFAR,
	author = {{van Haarlem, M. P.} and {Wise, M. W.} and {Gunst, A. W.} and {Heald, G.} and {McKean, J. P.} and {Hessels, J. W. T.} and {de Bruyn, A. G.} and {Nijboer, R.} and {Swinbank, J.} and {Fallows, R.} and {Brentjens, M.} and {Nelles, A.} and {Beck, R.} and {Falcke, H.} and {Fender, R.} and {Hörandel, J.} and {Koopmans, L. V. E.} and {Mann, G.} and {Miley, G.} and {Röttgering, H.} and {Stappers, B. W.} and {Wijers, R. A. M. J.} and {Zaroubi, S.} and {van den Akker, M.} and {Alexov, A.} and {Anderson, J.} and {Anderson, K.} and {van Ardenne, A.} and {Arts, M.} and {Asgekar, A.} and {Avruch, I. M.} and {Batejat, F.} and {Bähren, L.} and {Bell, M. E.} and {Bell, M. R.} and {van Bemmel, I.} and {Bennema, P.} and {Bentum, M. J.} and {Bernardi, G.} and {Best, P.} and {Bîrzan, L.} and {Bonafede, A.} and {Boonstra, A.-J.} and {Braun, R.} and {Bregman, J.} and {Breitling, F.} and {van de Brink, R. H.} and {Broderick, J.} and {Broekema, P. C.} and {Brouw, W. N.} and {Brüggen, M.} and {Butcher, H. R.} and {van Cappellen, W.} and {Ciardi, B.} and {Coenen, T.} and {Conway, J.} and {Coolen, A.} and {Corstanje, A.} and {Damstra, S.} and {Davies, O.} and {Deller, A. T.} and {Dettmar, R.-J.} and {van Diepen, G.} and {Dijkstra, K.} and {Donker, P.} and {Doorduin, A.} and {Dromer, J.} and {Drost, M.} and {van Duin, A.} and {Eislöffel, J.} and {van Enst, J.} and {Ferrari, C.} and {Frieswijk, W.} and {Gankema, H.} and {Garrett, M. A.} and {de Gasperin, F.} and {Gerbers, M.} and {de Geus, E.} and {Grießmeier, J.-M.} and {Grit, T.} and {Gruppen, P.} and {Hamaker, J. P.} and {Hassall, T.} and {Hoeft, M.} and {Holties, H. A.} and {Horneffer, A.} and {van der Horst, A.} and {van Houwelingen, A.} and {Huijgen, A.} and {Iacobelli, M.} and {Intema, H.} and {Jackson, N.} and {Jelic, V.} and {de Jong, A.} and {Juette, E.} and {Kant, D.} and {Karastergiou, A.} and {Koers, A.} and {Kollen, H.} and {Kondratiev, V. I.} and {Kooistra, E.} and {Koopman, Y.} and {Koster, A.} and {Kuniyoshi, M.} and {Kramer, M.} and {Kuper, G.} and {Lambropoulos, P.} and {Law, C.} and {van Leeuwen, J.} and {Lemaitre, J.} and {Loose, M.} and {Maat, P.} and {Macario, G.} and {Markoff, S.} and {Masters, J.} and {McFadden, R. A.} and {McKay-Bukowski, D.} and {Meijering, H.} and {Meulman, H.} and {Mevius, M.} and {Middelberg, E.} and {Millenaar, R.} and {Miller-Jones, J. C. A.} and {Mohan, R. N.} and {Mol, J. D.} and {Morawietz, J.} and {Morganti, R.} and {Mulcahy, D. D.} and {Mulder, E.} and {Munk, H.} and {Nieuwenhuis, L.} and {van Nieuwpoort, R.} and {Noordam, J. E.} and {Norden, M.} and {Noutsos, A.} and {Offringa, A. R.} and {Olofsson, H.} and {Omar, A.} and {Orrú, E.} and {Overeem, R.} and {Paas, H.} and {Pandey-Pommier, M.} and {Pandey, V. N.} and {Pizzo, R.} and {Polatidis, A.} and {Rafferty, D.} and {Rawlings, S.} and {Reich, W.} and {de Reijer, J.-P.} and {Reitsma, J.} and {Renting, G. A.} and {Riemers, P.} and {Rol, E.} and {Romein, J. W.} and {Roosjen, J.} and {Ruiter, M.} and {Scaife, A.} and {van der Schaaf, K.} and {Scheers, B.} and {Schellart, P.} and {Schoenmakers, A.} and {Schoonderbeek, G.} and {Serylak, M.} and {Shulevski, A.} and {Sluman, J.} and {Smirnov, O.} and {Sobey, C.} and {Spreeuw, H.} and {Steinmetz, M.} and {Sterks, C. G. M.} and {Stiepel, H.-J.} and {Stuurwold, K.} and {Tagger, M.} and {Tang, Y.} and {Tasse, C.} and {Thomas, I.} and {Thoudam, S.} and {Toribio, M. C.} and {van der Tol, B.} and {Usov, O.} and {van Veelen, M.} and {van der Veen, A.-J.} and {ter Veen, S.} and {Verbiest, J. P. W.} and {Vermeulen, R.} and {Vermaas, N.} and {Vocks, C.} and {Vogt, C.} and {de Vos, M.} and {van der Wal, E.} and {van Weeren, R.} and {Weggemans, H.} and {Weltevrede, P.} and {White, S.} and {Wijnholds, S. J.} and {Wilhelmsson, T.} and {Wucknitz, O.} and {Yatawatta, S.} and {Zarka, P.} and {Zensus, A.} and {van Zwieten, J.}},
	title = {LOFAR: The LOw-Frequency ARray},
	DOI= "10.1051/0004-6361/201220873",
	url= "https://doi.org/10.1051/0004-6361/201220873",
	journal = {A\&A},
	year = 2013,
	volume = 556,
	pages = "A2",
	month = "",
}

@article{DeBoer_2017,
doi = {10.1088/1538-3873/129/974/045001},
url = {https://dx.doi.org/10.1088/1538-3873/129/974/045001},
year = {2017},
month = {mar},
publisher = {The Astronomical Society of the Pacific},
volume = {129},
number = {974},
pages = {045001},
author = {David R. DeBoer and Aaron R. Parsons and James E. Aguirre and Paul Alexander and Zaki S. Ali and Adam P. Beardsley and Gianni Bernardi and Judd D. Bowman and Richard F. Bradley and Chris L. Carilli and Carina Cheng and Eloy de Lera Acedo and Joshua S. Dillon and Aaron Ewall-Wice and Gcobisa Fadana and Nicolas Fagnoni and Randall Fritz and Steve R. Furlanetto and Brian Glendenning and Bradley Greig and Jasper Grobbelaar and Bryna J. Hazelton and Jacqueline N. Hewitt and Jack Hickish and Daniel C. Jacobs and Austin Julius and MacCalvin Kariseb and Saul A. Kohn and Telalo Lekalake and Adrian Liu and Anita Loots and David MacMahon and Lourence Malan and Cresshim Malgas and Matthys Maree and Zachary Martinot and Nathan Mathison and Eunice Matsetela and Andrei Mesinger and Miguel F. Morales and Abraham R. Neben and Nipanjana Patra and Samantha Pieterse and Jonathan C. Pober and Nima Razavi-Ghods and Jon Ringuette and James Robnett and Kathryn Rosie and Raddwine Sell and Craig Smith and Angelo Syce and Max Tegmark and Nithyanandan Thyagarajan and Peter K. G. Williams and Haoxuan Zheng},
title = {Hydrogen Epoch of Reionization Array (HERA)},
journal = {PASP}
}

@article{Abdurashidova_2022,
   title={First Results from HERA Phase I: Upper Limits on the Epoch of Reionization 21 cm Power Spectrum},
   volume={925},
   ISSN={1538-4357},
   url={http://dx.doi.org/10.3847/1538-4357/ac1c78},
   DOI={10.3847/1538-4357/ac1c78},
   number={2},
   journal={ApJ},
   publisher={American Astronomical Society},
   author={Abdurashidova, Zara and Aguirre, James E. and Alexander, Paul and Ali, Zaki S. and Balfour, Yanga and Beardsley, Adam P. and Bernardi, Gianni and Billings, Tashalee S. and Bowman, Judd D. and Bradley, Richard F. and Bull, Philip and Burba, Jacob and Carey, Steve and Carilli, Chris L. and Cheng, Carina and DeBoer, David R. and Dexter, Matt and de Lera Acedo, Eloy and Dibblee-Barkman, Taylor and Dillon, Joshua S. and Ely, John and Ewall-Wice, Aaron and Fagnoni, Nicolas and Fritz, Randall and Furlanetto, Steven R. and Gale-Sides, Kingsley and Glendenning, Brian and Gorthi, Deepthi and Greig, Bradley and Grobbelaar, Jasper and Halday, Ziyaad and Hazelton, Bryna J. and Hewitt, Jacqueline N. and Hickish, Jack and Jacobs, Daniel C. and Julius, Austin and Kern, Nicholas S. and Kerrigan, Joshua and Kittiwisit, Piyanat and Kohn, Saul A. and Kolopanis, Matthew and Lanman, Adam and La Plante, Paul and Lekalake, Telalo and Lewis, David and Liu, Adrian and MacMahon, David and Malan, Lourence and Malgas, Cresshim and Maree, Matthys and Martinot, Zachary E. and Matsetela, Eunice and Mesinger, Andrei and Molewa, Mathakane and Morales, Miguel F. and Mosiane, Tshegofalang and Murray, Steven G. and Neben, Abraham R. and Nikolic, Bojan and Nunhokee, Chuneeta D. and Parsons, Aaron R. and Patra, Nipanjana and Pascua, Robert and Pieterse, Samantha and Pober, Jonathan C. and Razavi-Ghods, Nima and Ringuette, Jon and Robnett, James and Rosie, Kathryn and Sims, Peter and Singh, Saurabh and Smith, Craig and Syce, Angelo and Thyagarajan, Nithyanandan and Williams, Peter K. G. and Zheng, Haoxuan},
   year={2022},
   month=feb, pages={221} }

@article{Abdurashidova_2023,
   title={Improved Constraints on the 21 cm EoR Power Spectrum and the X-Ray Heating of the IGM with HERA Phase I Observations},
   volume={945},
   ISSN={1538-4357},
   url={http://dx.doi.org/10.3847/1538-4357/acaf50},
   DOI={10.3847/1538-4357/acaf50},
   number={2},
   journal={ApJ},
   publisher={American Astronomical Society},
   author={Abdurashidova, The HERA Collaboration: Zara and Adams, Tyrone and Aguirre, James E. and Alexander, Paul and Ali, Zaki S. and Baartman, Rushelle and Balfour, Yanga and Barkana, Rennan and Beardsley, Adam P. and Bernardi, Gianni and Billings, Tashalee S. and Bowman, Judd D. and Bradley, Richard F. and Breitman, Daniela and Bull, Philip and Burba, Jacob and Carey, Steve and Carilli, Chris L. and Cheng, Carina and Choudhuri, Samir and DeBoer, David R. and de Lera Acedo, Eloy and Dexter, Matt and Dillon, Joshua S. and Ely, John and Ewall-Wice, Aaron and Fagnoni, Nicolas and Fialkov, Anastasia and Fritz, Randall and Furlanetto, Steven R. and Gale-Sides, Kingsley and Garsden, Hugh and Glendenning, Brian and Gorce, Adélie and Gorthi, Deepthi and Greig, Bradley and Grobbelaar, Jasper and Halday, Ziyaad and Hazelton, Bryna J. and Heimersheim, Stefan and Hewitt, Jacqueline N. and Hickish, Jack and Jacobs, Daniel C. and Julius, Austin and Kern, Nicholas S. and Kerrigan, Joshua and Kittiwisit, Piyanat and Kohn, Saul A. and Kolopanis, Matthew and Lanman, Adam and La Plante, Paul and Lewis, David and Liu, Adrian and Loots, Anita and Ma, Yin-Zhe and MacMahon, David H. E. and Malan, Lourence and Malgas, Keith and Malgas, Cresshim and Maree, Matthys and Marero, Bradley and Martinot, Zachary E. and McBride, Lisa and Mesinger, Andrei and Mirocha, Jordan and Molewa, Mathakane and Morales, Miguel F. and Mosiane, Tshegofalang and Muñoz, Julian B. and Murray, Steven G. and Nagpal, Vighnesh and Neben, Abraham R. and Nikolic, Bojan and Nunhokee, Chuneeta D. and Nuwegeld, Hans and Parsons, Aaron R. and Pascua, Robert and Patra, Nipanjana and Pieterse, Samantha and Qin, Yuxiang and Razavi-Ghods, Nima and Robnett, James and Rosie, Kathryn and Santos, Mario G. and Sims, Peter and Singh, Saurabh and Smith, Craig and Swarts, Hilton and Tan, Jianrong and Thyagarajan, Nithyanandan and Wilensky, Michael J. and Williams, Peter K. G. and van Wyngaarden, Pieter and Zheng, Haoxuan},
   year={2023},
   month=mar, pages={124} }

@ARTICLE{Ceccotti_2025,
       author = {{Ceccotti}, E. and {Offringa}, A.~R. and {Mertens}, F.~G. and {Koopmans}, L.~V.~E. and {Munshi}, S. and {Chege}, J.~K. and {Acharya}, A. and {Brackenhoff}, S.~A. and {Chapman}, E. and {Ciardi}, B. and {Ghara}, R. and {Ghosh}, S. and {Giri}, S.~K. and {H{\"o}fer}, C. and {Hothi}, I. and {Mellema}, G. and {Mevius}, M. and {Pandey}, V.~N. and {Zaroubi}, S.},
        title = "{First upper limits on the 21-cm signal power spectrum of neutral hydrogen at z = 9.16 from the LOFAR 3C 196 field}",
      journal = {\mnras},
         year = 2025,
        month = nov,
       volume = {544},
       number = {1},
        pages = {1255-1283},
          doi = {10.1093/mnras/staf1629},
archivePrefix = {arXiv},
       eprint = {2504.18534},
 primaryClass = {astro-ph.CO},
       adsurl = {https://ui.adsabs.harvard.edu/abs/2025MNRAS.544.1255C}
}

@ARTICLE{Trott_2020,
       author = {{Trott}, Cathryn M. and {Jordan}, C.~H. and {Midgley}, S. and {Barry}, N. and {Greig}, B. and {Pindor}, B. and {Cook}, J.~H. and {Sleap}, G. and {Tingay}, S.~J. and {Ung}, D. and {Hancock}, P. and {Williams}, A. and {Bowman}, J. and {Byrne}, R. and {Chokshi}, A. and {Hazelton}, B.~J. and {Hasegawa}, K. and {Jacobs}, D. and {Joseph}, R.~C. and {Li}, W. and {Line}, J.~L.~B. and {Lynch}, C. and {McKinley}, B. and {Mitchell}, D.~A. and {Morales}, M.~F. and {Ouchi}, M. and {Pober}, J.~C. and {Rahimi}, M. and {Takahashi}, K. and {Wayth}, R.~B. and {Webster}, R.~L. and {Wilensky}, M. and {Wyithe}, J.~S.~B. and {Yoshiura}, S. and {Zhang}, Z. and {Zheng}, Q.},
        title = "{Deep multiredshift limits on Epoch of Reionization 21 cm power spectra from four seasons of Murchison Widefield Array observations}",
      journal = {\mnras},
         year = 2020,
        month = apr,
       volume = {493},
       number = {4},
        pages = {4711-4727},
          doi = {10.1093/mnras/staa414},
archivePrefix = {arXiv},
       eprint = {2002.02575},
 primaryClass = {astro-ph.CO},
       adsurl = {https://ui.adsabs.harvard.edu/abs/2020MNRAS.493.4711T}
}

@ARTICLE{Mertens_2020,
       author = {{Mertens}, F.~G. and {Mevius}, M. and {Koopmans}, L.~V.~E. and {Offringa}, A.~R. and {Mellema}, G. and {Zaroubi}, S. and {Brentjens}, M.~A. and {Gan}, H. and {Gehlot}, B.~K. and {Pandey}, V.~N. and {Sardarabadi}, A.~M. and {Vedantham}, H.~K. and {Yatawatta}, S. and {Asad}, K.~M.~B. and {Ciardi}, B. and {Chapman}, E. and {Gazagnes}, S. and {Ghara}, R. and {Ghosh}, A. and {Giri}, S.~K. and {Iliev}, I.~T. and {Jeli{\'c}}, V. and {Kooistra}, R. and {Mondal}, R. and {Schaye}, J. and {Silva}, M.~B.},
        title = "{Improved upper limits on the 21 cm signal power spectrum of neutral hydrogen at z {\ensuremath{\approx}} 9.1 from LOFAR}",
      journal = {\mnras},
         year = 2020,
        month = apr,
       volume = {493},
       number = {2},
        pages = {1662-1685},
          doi = {10.1093/mnras/staa327},
archivePrefix = {arXiv},
       eprint = {2002.07196},
 primaryClass = {astro-ph.CO},
       adsurl = {https://ui.adsabs.harvard.edu/abs/2020MNRAS.493.1662M}
}

@ARTICLE{Yoshiura_2021,
       author = {{Yoshiura}, S. and {Pindor}, B. and {Line}, J.~L.~B. and {Barry}, N. and {Trott}, C.~M. and {Beardsley}, A. and {Bowman}, J. and {Byrne}, R. and {Chokshi}, A. and {Hazelton}, B.~J. and {Hasegawa}, K. and {Howard}, E. and {Greig}, B. and {Jacobs}, D. and {Jordan}, C.~H. and {Joseph}, R. and {Kolopanis}, M. and {Lynch}, C. and {McKinley}, B. and {Mitchell}, D.~A. and {Morales}, M.~F. and {Murray}, S.~G. and {Pober}, J.~C. and {Rahimi}, M. and {Takahashi}, K. and {Tingay}, S.~J. and {Wayth}, R.~B. and {Webster}, R.~L. and {Wilensky}, M. and {Wyithe}, J.~S.~B. and {Zhang}, Z. and {Zheng}, Q.},
        title = "{A new MWA limit on the 21 cm power spectrum at redshifts  13-17}",
      journal = {\mnras},
         year = 2021,
        month = aug,
       volume = {505},
       number = {4},
        pages = {4775-4790},
          doi = {10.1093/mnras/stab1560},
archivePrefix = {arXiv},
       eprint = {2105.12888},
 primaryClass = {astro-ph.CO},
       adsurl = {https://ui.adsabs.harvard.edu/abs/2021MNRAS.505.4775Y}
}

@Article{Schosser_2025,
	title={{Optimal, fast, and robust inference of reionization-era cosmology with the 21cmPIE-INN}},
	author={Benedikt Schosser and Caroline Heneka and Tilman Plehn},
	journal={SciPost Phys. Core},
	volume={8},
	pages={037},
	year={2025},
	publisher={SciPost},
	doi={10.21468/SciPostPhysCore.8.2.037},
	url={https://scipost.org/10.21468/SciPostPhysCore.8.2.037},
}

@Article{Ore_2025,
	title={{SKATR: A self-supervised summary transformer for SKA}},
	author={Ayodele Ore and Caroline Heneka and Tilman Plehn},
	journal={SciPost Phys.},
	volume={18},
	pages={155},
	year={2025},
	publisher={SciPost},
	doi={10.21468/SciPostPhys.18.5.155},
	url={https://scipost.org/10.21468/SciPostPhys.18.5.155},
}

@article{Park_2019,
   title={Inferring the astrophysics of reionization and cosmic dawn from galaxy luminosity functions and the 21-cm signal},
   volume={484},
   ISSN={1365-2966},
   url={http://dx.doi.org/10.1093/mnras/stz032},
   DOI={10.1093/mnras/stz032},
   number={1},
   journal={\mnras},
   publisher={Oxford University Press (OUP)},
   author={Park, Jaehong and Mesinger, Andrei and Greig, Bradley and Gillet, Nicolas},
   year={2019},
   month=jan, pages={933–949} }

@article{21cmfast, doi = {10.21105/joss.02582}, url = {https://doi.org/10.21105/joss.02582}, year = {2020}, publisher = {The Open Journal}, volume = {5}, number = {54}, pages = {2582}, author = {Steven G. Murray and Bradley Greig and Andrei Mesinger and Julian B. Muñoz and Yuxiang Qin and Jaehong Park and Catherine A. Watkinson}, title = {21cmFAST v3: A Python-integrated C code for generating 3D realizations of the cosmic 21cm signal.}, journal = {JOSS} }

@article{21cmfast11,
	title = {21cmfast: a fast, seminumerical simulation of the high-redshift 21-cm signal},
	volume = {411},
	issn = {0035-8711},
	url = {https://doi.org/10.1111/j.1365-2966.2010.17731.x},
	doi = {10.1111/j.1365-2966.2010.17731.x},
	number = {2},
	journal = {\mnras},
	author = {Mesinger, Andrei and Furlanetto, Steven and Cen, Renyue},
	month = feb,
	year = {2011},
	pages = {955--972},
}

@article{21cmsense13,
doi = {10.1088/0004-6256/145/3/65},
url = {https://dx.doi.org/10.1088/0004-6256/145/3/65},
year = {2013},
month = {jan},
publisher = {The American Astronomical Society},
volume = {145},
number = {3},
pages = {65},
author = {Jonathan C. Pober and Aaron R. Parsons and David R. DeBoer and Patrick McDonald and Matthew McQuinn and James E. Aguirre and Zaki Ali and Richard F. Bradley and Tzu-Ching Chang and Miguel F. Morales},
title = {THE BARYON ACOUSTIC OSCILLATION BROADBAND AND BROAD-BEAM ARRAY: DESIGN OVERVIEW AND SENSITIVITY FORECASTS},
journal = {\aj}
}

@article{21cmsense14,
doi = {10.1088/0004-637X/782/2/66},
url = {https://dx.doi.org/10.1088/0004-637X/782/2/66},
year = {2014},
month = {jan},
publisher = {The American Astronomical Society},
volume = {782},
number = {2},
pages = {66},
author = {Jonathan C. Pober and Adrian Liu and Joshua S. Dillon and James E. Aguirre and Judd D. Bowman and Richard F. Bradley and Chris L. Carilli and David R. DeBoer and Jacqueline N. Hewitt and Daniel C. Jacobs and Matthew McQuinn and Miguel F. Morales and Aaron R. Parsons and Max Tegmark and Dan J. Werthimer},
title = {WHAT NEXT-GENERATION 21 cm POWER SPECTRUM MEASUREMENTS CAN TEACH US ABOUT THE EPOCH OF REIONIZATION},
journal = {ApJ}
}

@inproceedings{Durkan_Spline,
 author = {Durkan, Conor and Bekasov, Artur and Murray, Iain and Papamakarios, George},
 booktitle = {Proceedings of NeurIPS},
 publisher = {Curran Associates, Inc.},
 title = {Neural Spline Flows},
 url = {https://proceedings.neurips.cc/paper_files/paper/2019/file/7ac71d433f282034e088473244df8c02-Paper.pdf},
 volume = {32},
 year = {2019}
}

@inproceedings{Pytorch,
author = {Ansel, Jason and Yang, Edward and He, Horace and Gimelshein, Natalia and Jain, Animesh and Voznesensky, Michael and Bao, Bin and Bell, Peter and Berard, David and Burovski, Evgeni and Chauhan, Geeta and Chourdia, Anjali and Constable, Will and Desmaison, Alban and DeVito, Zachary and Ellison, Elias and Feng, Will and Gong, Jiong and Gschwind, Michael and Hirsh, Brian and Huang, Sherlock and Kalambarkar, Kshiteej and Kirsch, Laurent and Lazos, Michael and Lezcano, Mario and Liang, Yanbo and Liang, Jason and Lu, Yinghai and Luk, CK and Maher, Bert and Pan, Yunjie and Puhrsch, Christian and Reso, Matthias and Saroufim, Mark and Siraichi, Marcos Yukio and Suk, Helen and Suo, Michael and Tillet, Phil and Wang, Eikan and Wang, Xiaodong and Wen, William and Zhang, Shunting and Zhao, Xu and Zhou, Keren and Zou, Richard and Mathews, Ajit and Chanan, Gregory and Wu, Peng and Chintala, Soumith},
booktitle = {Proceedings of ASPLOS},
doi = {10.1145/3620665.3640366},
month = apr,
publisher = {ACM},
title = {{PyTorch 2: Faster Machine Learning Through Dynamic Python Bytecode Transformation and Graph Compilation}},
url = {https://pytorch.org/assets/pytorch2-2.pdf},
year = {2024}
}

@inproceedings{
AdamW,
title={Decoupled Weight Decay Regularization},
author={Ilya Loshchilov and Frank Hutter},
booktitle={Proceedings of ICLR},
year={2019},
url={https://openreview.net/forum?id=Bkg6RiCqY7},
}

@ARTICLE{Pietschke_2025,
       author = {{Pietschke}, Yannic and {Heneka}, Caroline and {Schlenker}, Tom and {Ore}, Ayodele and {Schosser}, Benedikt},
        title = "{Direct reconstruction of the Reionization history from 21cm 2D Power Spectra}",
      journal = {JCAP},
         year = 2025,
        month = oct,
       volume = {2025},
       number = {10},
          eid = {039},
        pages = {039},
          doi = {10.1088/1475-7516/2025/10/039},
archivePrefix = {arXiv},
       eprint = {2506.19925},
 primaryClass = {astro-ph.CO},
       adsurl = {https://ui.adsabs.harvard.edu/abs/2025JCAP...10..039P}
}

@ARTICLE{Hutter_2025,
       author = {{Hutter}, Anne and {Heneka}, Caroline},
        title = "{The 21cm─galaxy cross-correlation: Realistic forecast for 21cm signal detection and reionisation constraints}",
      journal = {\aap},
         year = 2026,
        month = mar,
       volume = {707},
          eid = {A286},
        pages = {A286},
          doi = {10.1051/0004-6361/202557324},
archivePrefix = {arXiv},
       eprint = {2509.15906},
 primaryClass = {astro-ph.CO},
       adsurl = {https://ui.adsabs.harvard.edu/abs/2026A&A...707A.286H}
}

@ARTICLE{Davies_2025,
       author = {{Davies}, James E. and {Mesinger}, Andrei and {Murray}, Steven G.},
        title = "{Efficient simulation of discrete galaxy populations and associated radiation fields over the first billion years}",
      journal = {\aap},
         year = 2025,
        month = sep,
       volume = {701},
          eid = {A236},
        pages = {A236},
          doi = {10.1051/0004-6361/202554951},
archivePrefix = {arXiv},
       eprint = {2504.17254},
 primaryClass = {astro-ph.CO},
       adsurl = {https://ui.adsabs.harvard.edu/abs/2025A&A...701A.236D}
}

@ARTICLE{Cerardi_2025,
       author = {{Cerardi}, Nicolas and {Giri}, Sambit K. and {Bianco}, Michele and {Piras}, Davide and {de Salis}, Emmanuel and {De Santis}, Massimo and {Selcuk-Simsek}, Merve and {Denzel}, Philipp and {Hess}, Kelley M. and {Toribio}, M. Carmen and {Kirsten}, Franz and {Ghorbel}, Hatem},
        title = "{Implicit inference of the reionization history with higher-order statistics of the 21-cm signal}",
      journal = {Submitted to MNRAS},
         year = 2025,
        month = nov,
          eid = {arXiv:2511.11568},
        pages = {arXiv:2511.11568},
          doi = {10.48550/arXiv.2511.11568},
archivePrefix = {arXiv},
       eprint = {2511.11568},
 primaryClass = {astro-ph.CO},
       adsurl = {https://ui.adsabs.harvard.edu/abs/2025arXiv251111568C}
}

@ARTICLE{Cooper_2025,
    author = {Cooper, Nadia and Norregaard, Carina and Meriot, Romain and Pritchard, Jonathan R},
    title = {Simulation-based inference pipeline of the ionization history from the 2D 21 cm power spectrum},
    journal = {\mnras},
    volume = {548},
    number = {2},
    pages = {stag577},
    year = {2026},
    month = {05},
    issn = {0035-8711},
    doi = {10.1093/mnras/stag577},
    url = {https://doi.org/10.1093/mnras/stag577},
}

@ARTICLE{Beane2019,
       author = {{Beane}, Angus and {Villaescusa-Navarro}, Francisco and {Lidz}, Adam},
        title = "{Measuring the EoR Power Spectrum without Measuring the EoR Power Spectrum}",
      journal = {\apj},
         year = 2019,
        month = apr,
       volume = {874},
       number = {2},
          eid = {133},
        pages = {133},
          doi = {10.3847/1538-4357/ab0a08},
archivePrefix = {arXiv},
       eprint = {1811.10609},
 primaryClass = {astro-ph.CO},
       adsurl = {https://ui.adsabs.harvard.edu/abs/2019ApJ...874..133B}
}

@ARTICLE{HERA2025,
       author = {{Abdurashidova}, Zuhra and {Adams}, Tyrone and {Aguirre}, James E. and {Baartman}, Rushelle and {Barkana}, Rennan and {Berkhout}, Lindsay M. and {Bernardi}, Gianni and {Billings}, Tashalee S. and {Bizarria}, Bruno B. and {Bowman}, Judd D. and {Breitman}, Daniela and {Bull}, Philip and {Burba}, Jacob and {Byrne}, Ruby and {Carey}, Steven and {Chandra}, Rajorshi Sushovan and {Chen}, Kai-Feng and {Choudhuri}, Samir and {Cox}, Tyler and {DeBoer}, David R. and {de Lera Acedo}, Eloy and {Dexter}, Matt and {Dhandha}, Jiten and {Dillon}, Joshua S. and {Dynes}, Scott and {Eksteen}, Nico and {Ely}, John and {Ewall-Wice}, Aaron and {Fagnoni}, Nicolas and {Fialkov}, Anastasia and {Furlanetto}, Steven R. and {Gale-Sides}, Kingsley and {Garsden}, Hugh and {Gorce}, Adelie and {Gorthi}, Deepthi and {Halday}, Ziyaad and {Hazelton}, Bryna J. and {Hewitt}, Jacqueline N. and {Hickish}, Jack and {Huang}, Tian and {Jacobs}, Daniel C. and {Josaitis}, Alec and {Kern}, Nicholas S. and {Kerrigan}, Joshua and {Kittiwisit}, Piyanat and {Kolopanis}, Matthew and {Lanman}, Adam and {La Plante}, Paul and {Liu}, Adrian and {Ma}, Yin-Zhe and {MacMahon}, David H.~E. and {Malan}, Lourence and {Malgas}, Cresshim and {Malgas}, Keith and {Marero}, Bradley and {Martinot}, Zachary E. and {McBride}, Lisa and {Mesinger}, Andrei and {Mirocha}, Jordan and {Mohamed-Hinds}, Nicel and {Molewa}, Mathakane and {Morales}, Miguel F. and {Mu{\~n}oz}, Julian B. and {Murray}, Steven G. and {Nikolic}, Bojan and {Nuwegeld}, Hans and {Parsons}, Aaron R. and {Pascua}, Robert and {Patra}, Nipanjana and {Pochinda}, Simon and {Qin}, Yuxiang and {Rath}, Eleanor and {Razavi-Ghods}, Nima and {Riley}, Daniel and {Rosie}, Kathryn and {Santos}, Mario G. and {Singh}, Saurabh and {Storer}, Dara and {Swarts}, Hilton and {Tan}, Jianrong and {Th{\'e}lie}, Emilie and {van Wyngaarden}, Pieter and {Wilensky}, Michael J. and {Williams}, Peter K.~G. and {Zheng}, Haoxuan and {HERA Collaboration}},
        title = "{First Results from HERA Phase II}",
      journal = {\apj},
         year = 2026,
        month = feb,
       volume = {998},
       number = {1},
          eid = {33},
        pages = {33},
          doi = {10.3847/1538-4357/ae2d54},
archivePrefix = {arXiv},
       eprint = {2511.21289},
 primaryClass = {astro-ph.CO},
       adsurl = {https://ui.adsabs.harvard.edu/abs/2026ApJ...998...33A}
}

@ARTICLE{acharya2024,
       author = {{Acharya}, Anshuman and {Mertens}, Florent and {Ciardi}, Benedetta and {Ghara}, Raghunath and {Koopmans}, L{\'e}on V.~E. and {Zaroubi}, Saleem},
        title = "{Revised LOFAR upper limits on the 21-cm signal power spectrum at z {\ensuremath{\approx}} 9.1 using machine learning and gaussian process regression}",
      journal = {\mnras},
         year = 2024,
        month = oct,
       volume = {534},
       number = {1},
        pages = {L30-L34},
          doi = {10.1093/mnrasl/slae078},
archivePrefix = {arXiv},
       eprint = {2408.10051},
 primaryClass = {astro-ph.CO},
       adsurl = {https://ui.adsabs.harvard.edu/abs/2024MNRAS.534L..30A}
}

@ARTICLE{Moriwaki2024,
       author = {{Moriwaki}, Kana and {Beane}, Angus and {Lidz}, Adam},
        title = "{Insights into the 21 cm field from the vanishing cross-power spectrum at the epoch of reionization}",
      journal = {\mnras},
         year = 2024,
        month = may,
       volume = {530},
       number = {3},
        pages = {3183-3194},
          doi = {10.1093/mnras/stae1050},
archivePrefix = {arXiv},
       eprint = {2404.08266},
 primaryClass = {astro-ph.CO},
       adsurl = {https://ui.adsabs.harvard.edu/abs/2024MNRAS.530.3183M}
}

@ARTICLE{Sobacchi2016,
       author = {{Sobacchi}, Emanuele and {Mesinger}, Andrei and {Greig}, Bradley},
        title = "{Cross-correlation of the cosmic 21-cm signal and Lyman {\ensuremath{\alpha}} emitters during reionization}",
      journal = {\mnras},
         year = 2016,
        month = jul,
       volume = {459},
       number = {3},
        pages = {2741-2750},
          doi = {10.1093/mnras/stw811},
archivePrefix = {arXiv},
       eprint = {1602.04837},
 primaryClass = {astro-ph.CO},
       adsurl = {https://ui.adsabs.harvard.edu/abs/2016MNRAS.459.2741S}
}

@ARTICLE{Yoshiura2018,
       author = {{Yoshiura}, S. and {Line}, J.~L.~B. and {Kubota}, K. and {Hasegawa}, K. and {Takahashi}, K.},
        title = "{Detectability of 21 cm-signal during the Epoch of Reionization with 21 cm-Lyman-{\ensuremath{\alpha}} emitter cross-correlation - II. Foreground contamination}",
      journal = {\mnras},
         year = 2018,
        month = sep,
       volume = {479},
       number = {2},
        pages = {2767-2776},
          doi = {10.1093/mnras/sty1472},
archivePrefix = {arXiv},
       eprint = {1709.04168},
 primaryClass = {astro-ph.CO},
       adsurl = {https://ui.adsabs.harvard.edu/abs/2018MNRAS.479.2767Y}
}

@MISC{Greene2022,
       author = {{Greene}, Jenny and {Bezanson}, Rachel and {Ouchi}, Masami and {Silverman}, John and {the PFS Galaxy Evolution Working Group}},
        title = "{The Prime Focus Spectrograph Galaxy Evolution Survey}",
      journal = {arXiv e-prints},
         year = 2022,
        month = jun,
          eid = {arXiv:2206.14908},
        pages = {arXiv:2206.14908},
          doi = {10.48550/arXiv.2206.14908},
archivePrefix = {arXiv},
       eprint = {2206.14908},
 primaryClass = {astro-ph.GA},
       adsurl = {https://ui.adsabs.harvard.edu/abs/2022arXiv220614908G}
}

@ARTICLE{Maiolino2020,
       author = {{Maiolino}, R. and {Cirasuolo}, M. and {Afonso}, J. and {Bauer}, F.~E. and {Bowler}, R. and {Cucciati}, O. and {Daddi}, E. and {De Lucia}, G. and {Evans}, C. and {Flores}, H. and {Gargiulo}, A. and {Garilli}, B. and {Jablonka}, P. and {Jarvis}, M. and {Kneib}, J. -P. and {Lilly}, S. and {Looser}, T. and {Magliocchetti}, M. and {Man}, Z. and {Mannucci}, F. and {Maurogordato}, S. and {McLure}, R.~J. and {Norberg}, P. and {Oesch}, P. and {Oliva}, E. and {Paltani}, S. and {Pappalardo}, C. and {Peng}, Y. and {Pentericci}, L. and {Pozzetti}, L. and {Renzini}, A. and {Rodrigues}, M. and {Royer}, F. and {Serjeant}, S. and {Vanzi}, L. and {Wild}, V. and {Zamorani}, G.},
        title = "{MOONRISE: The Main MOONS GTO Extragalactic Survey}",
      journal = {The Messenger},
         year = 2020,
        month = jun,
       volume = {180},
        pages = {24-29},
          doi = {10.18727/0722-6691/5197},
archivePrefix = {arXiv},
       eprint = {2009.00644},
 primaryClass = {astro-ph.GA},
       adsurl = {https://ui.adsabs.harvard.edu/abs/2020Msngr.180...24M}
}

@ARTICLE{Hutter2017,
       author = {{Hutter}, Anne and {Dayal}, Pratika and {M{\"u}ller}, Volker and {Trott}, Cathryn M.},
        title = "{Exploring 21cm-Lyman Alpha Emitter Synergies for SKA}",
      journal = {\apj},
         year = 2017,
        month = feb,
       volume = {836},
       number = {2},
          eid = {176},
        pages = {176},
          doi = {10.3847/1538-4357/836/2/176},
archivePrefix = {arXiv},
       eprint = {1605.01734},
 primaryClass = {astro-ph.CO},
       adsurl = {https://ui.adsabs.harvard.edu/abs/2017ApJ...836..176H}
}

@ARTICLE{Hutter2023a,
       author = {{Hutter}, Anne and {Trebitsch}, Maxime and {Dayal}, Pratika and {Gottl{\"o}ber}, Stefan and {Yepes}, Gustavo and {Legrand}, Laurent},
        title = "{ASTRAEUS - VIII. A new framework for Lyman-{\ensuremath{\alpha}} emitters applied to different reionization scenarios}",
      journal = {\mnras},
         year = 2023,
        month = oct,
       volume = {524},
       number = {4},
        pages = {6124-6148},
          doi = {10.1093/mnras/stad2230},
archivePrefix = {arXiv},
       eprint = {2209.14592},
 primaryClass = {astro-ph.GA},
       adsurl = {https://ui.adsabs.harvard.edu/abs/2023MNRAS.524.6124H}
}

@ARTICLE{Hutter2023b,
       author = {{Hutter}, Anne and {Heneka}, Caroline and {Dayal}, Pratika and {Gottl{\"o}ber}, Stefan and {Mesinger}, Andrei and {Trebitsch}, Maxime and {Yepes}, Gustavo},
        title = "{On the general nature of 21-cm-Lyman {\ensuremath{\alpha}} emitter cross-correlations during reionization}",
      journal = {\mnras},
         year = 2023,
        month = oct,
       volume = {525},
       number = {2},
        pages = {1664-1676},
          doi = {10.1093/mnras/stad2376},
archivePrefix = {arXiv},
       eprint = {2306.03156},
 primaryClass = {astro-ph.CO},
       adsurl = {https://ui.adsabs.harvard.edu/abs/2023MNRAS.525.1664H}
}

@ARTICLE{Villaescusa-Navarro_2022,
       author = {{Villaescusa-Navarro}, Francisco and {Genel}, Shy and {Angl{\'e}s-Alc{\'a}zar}, Daniel and {Thiele}, Leander and {Dave}, Romeel and {Narayanan}, Desika and {Nicola}, Andrina and {Li}, Yin and {Villanueva-Domingo}, Pablo and {Wandelt}, Benjamin and {Spergel}, David N. and {Somerville}, Rachel S. and {Zorrilla Matilla}, Jose Manuel and {Mohammad}, Faizan G. and {Hassan}, Sultan and {Shao}, Helen and {Wadekar}, Digvijay and {Eickenberg}, Michael and {Wong}, Kaze W.~K. and {Contardo}, Gabriella and {Jo}, Yongseok and {Moser}, Emily and {Lau}, Erwin T. and {Machado Poletti Valle}, Luis Fernando and {Perez}, Lucia A. and {Nagai}, Daisuke and {Battaglia}, Nicholas and {Vogelsberger}, Mark},
        title = "{The CAMELS Multifield Data Set: Learning the Universe's Fundamental Parameters with Artificial Intelligence}",
      journal = {\apjs},
         year = 2022,
        month = apr,
       volume = {259},
       number = {2},
          eid = {61},
        pages = {61},
          doi = {10.3847/1538-4365/ac5ab0},
archivePrefix = {arXiv},
       eprint = {2109.10915},
 primaryClass = {cs.LG},
       adsurl = {https://ui.adsabs.harvard.edu/abs/2022ApJS..259...61V}
}

@ARTICLE{Alsing_2019,
       author = {{Alsing}, Justin and {Charnock}, Tom and {Feeney}, Stephen and {Wandelt}, Benjamin},
        title = "{Fast likelihood-free cosmology with neural density estimators and active learning}",
      journal = {\mnras},
         year = 2019,
        month = sep,
       volume = {488},
       number = {3},
        pages = {4440-4458},
          doi = {10.1093/mnras/stz1960},
archivePrefix = {arXiv},
       eprint = {1903.00007},
 primaryClass = {astro-ph.CO},
       adsurl = {https://ui.adsabs.harvard.edu/abs/2019MNRAS.488.4440A}
}

@ARTICLE{Cole_2022,
       author = {{Cole}, Alex and {Miller}, Benjamin K. and {Witte}, Samuel J. and {Cai}, Maxwell X. and {Grootes}, Meiert W. and {Nattino}, Francesco and {Weniger}, Christoph},
        title = "{Fast and credible likelihood-free cosmology with truncated marginal neural ratio estimation}",
      journal = {JCAP},
         year = 2022,
        month = sep,
       volume = {2022},
       number = {9},
          eid = {004},
        pages = {004},
          doi = {10.1088/1475-7516/2022/09/004},
archivePrefix = {arXiv},
       eprint = {2111.08030},
 primaryClass = {astro-ph.CO},
       adsurl = {https://ui.adsabs.harvard.edu/abs/2022JCAP...09..004C}
}

@ARTICLE{Saxena_2024,
       author = {{Saxena}, Anchal and {Meerburg}, P. Daniel and {Weniger}, Christoph and {Acedo}, Eloy de Lera and {Handley}, Will},
        title = "{Simulation-based inference of the sky-averaged 21-cm signal from CD-EoR with REACH}",
      journal = {RASTI},
         year = 2024,
        month = jan,
       volume = {3},
       number = {1},
        pages = {724-736},
          doi = {10.1093/rasti/rzae047},
archivePrefix = {arXiv},
       eprint = {2403.14618},
 primaryClass = {astro-ph.CO},
       adsurl = {https://ui.adsabs.harvard.edu/abs/2024RASTI...3..724S}
}

@MISC{WST_2024,
       author = {{Mainieri}, Vincenzo and {Anderson}, Richard I. and {Brinchmann}, Jarle and {Cimatti}, Andrea and {Ellis}, Richard S. and {Hill}, Vanessa and {Kneib}, Jean-Paul and {McLeod}, Anna F. and {Opitom}, Cyrielle and {Roth}, Martin M. and {Sanchez-Saez}, Paula and {Smiljanic}, Rodolfo and {Tolstoy}, Eline and {Bacon}, Roland and {Randich}, Sofia and {Adamo}, Angela and {Annibali}, Francesca and {Arevalo}, Patricia and {Audard}, Marc and {Barsanti}, Stefania and {Battaglia}, Giuseppina and {Bayo Aran}, Amelia M. and {Belfiore}, Francesco and {Bellazzini}, Michele and {Bellini}, Emilio and {Beltran}, Maria Teresa and {Berni}, Leda and {Bianchi}, Simone and {Biazzo}, Katia and {Bisero}, Sofia and {Bisogni}, Susanna and {Bland-Hawthorn}, Joss and {Blondin}, Stephane and {Bodensteiner}, Julia and {Boffin}, Henri M.~J. and {Bonito}, Rosaria and {Bono}, Giuseppe and {Bouche}, Nicolas F. and {Bowman}, Dominic and {Braga}, Vittorio F. and {Bragaglia}, Angela and {Branchesi}, Marica and {Brucalassi}, Anna and {Bryant}, Julia J. and {Bryson}, Ian and {Busa}, Innocenza and {Camera}, Stefano and {Carbone}, Carmelita and {Casali}, Giada and {Casali}, Mark and {Casasola}, Viviana and {Castro}, Norberto and {Catelan}, Marcio and {Cavallo}, Lorenzo and {Chiappini}, Cristina and {Cioni}, Maria-Rosa and {Colless}, Matthew and {Colzi}, Laura and {Contarini}, Sofia and {Couch}, Warrick and {D'Ammando}, Filippo and {d'Assignies D.}, William and {D'Orazi}, Valentina and {da Silva}, Ronaldo and {Dainotti}, Maria Giovanna and {Damiani}, Francesco and {Danielski}, Camilla and {De Cia}, Annalisa and {de Jong}, Roelof S. and {Dhawan}, Suhail and {Dierickx}, Philippe and {Driver}, Simon P. and {Dupletsa}, Ulyana and {Escoffier}, Stephanie and {Escorza}, Ana and {Fabrizio}, Michele and {Fiorentino}, Giuliana and {Fontana}, Adriano and {Fontani}, Francesco and {Forero Sanchez}, Daniel and {Franois}, Patrick and {Galindo-Guil}, Francisco Jose and {Gallazzi}, Anna Rita and {Galli}, Daniele and {Garcia}, Miriam and {Garcia-Rojas}, Jorge and {Garilli}, Bianca and {Grand}, Robert and {Guarcello}, Mario Giuseppe and {Hazra}, Nandini and {Helmi}, Amina and {Herrero}, Artemio and {Iglesias}, Daniela and {Ilic}, Dragana and {Irsic}, Vid and {Ivanov}, Valentin D. and {Izzo}, Luca and {Jablonka}, Pascale and {Joachimi}, Benjamin and {Kakkad}, Darshan and {Kamann}, Sebastian and {Koposov}, Sergey and {Kordopatis}, Georges and {Kovacevic}, Andjelka B. and {Kraljic}, Katarina and {Kuncarayakti}, Hanindyo and {Kwon}, Yuna and {La Forgia}, Fiorangela and {Lahav}, Ofer and {Laigle}, Clotilde and {Lazzarin}, Monica and {Leaman}, Ryan and {Leclercq}, Floriane and {Lee}, Khee-Gan and {Lee}, David and {Lehnert}, Matt D. and {Lira}, Paulina and {Loffredo}, Eleonora and {Lucatello}, Sara and {Magrini}, Laura and {Maguire}, Kate and {Mahler}, Guillaume and {Zahra Majidi}, Fatemeh and {Malavasi}, Nicola and {Mannucci}, Filippo and {Marconi}, Marcella and {Martin}, Nicolas and {Marulli}, Federico and {Massari}, Davide and {Matsuno}, Tadafumi and {Mattheee}, Jorryt and {McGee}, Sean and {Merc}, Jaroslav and {Merle}, Thibault and {Miglio}, Andrea and {Migliorini}, Alessandra and {Minchev}, Ivan and {Minniti}, Dante and {Miret-Roig}, Nuria and {Monreal Ibero}, Ana and {Montano}, Federico and {Montet}, Ben T. and {Moresco}, Michele and {Moretti}, Chiara and {Moscardini}, Lauro and {Moya}, Andres and {Mueller}, Oliver and {Nanayakkara}, Themiya and {Nicholl}, Matt and {Nordlander}, Thomas and {Onori}, Francesca and {Padovani}, Marco and {Pala}, Anna Francesca and {Panda}, Swayamtrupta and {Pandey-Pommier}, Mamta and {Pasquini}, Luca and {Pawlak}, Michal and {Pessi}, Priscila J. and {Pisani}, Alice and {Popovic}, Lukav C. and {Prisinzano}, Loredana and {Raddi}, Roberto and {Rainer}, Monica and {Rebassa-Mansergas}, Alberto and {Richard}, Johan and {Rigault}, Mickael and {Rocher}, Antoine and {Romano}, Donatella and {Rosati}, Piero and {Sacco}, Germano and {Sanchez-Janssen}, Ruben and {Sander}, Andreas A.~C. and {Sanders}, Jason L. and {Sargent}, Mark and {Sarpa}, Elena and {Schimd}, Carlo and {Schipani}, Pietro and {Sefusatti}, Emiliano and {Smith}, Graham P. and {Spina}, Lorenzo and {Steinmetz}, Matthias and {Tacchella}, Sandro and {Tautvaisiene}, Grazina and {Theissen}, Christopher and {Thomas}, Guillaume and {Ting}, Yuan-Sen and {Travouillon}, Tony and {Tresse}, Laurence and {Trivedi}, Oem and {Tsantaki}, Maria and {Tsedrik}, Maria and {Urrutia}, Tanya and {Valenti}, Elena and {Van der Swaelmen}, Mathieu and {Van Eck}, Sophie and {Verdiani}, Francesco and {Verdier}, Aurelien and {Vergani}, Susanna Diana and {Verhamme}, Anne and {Vernet}, Joel},
        title = "{The Wide-field Spectroscopic Telescope (WST) Science White Paper}",
      journal = {arXiv e-prints},
         year = 2024,
        month = mar,
          eid = {arXiv:2403.05398},
        pages = {arXiv:2403.05398},
          doi = {10.48550/arXiv.2403.05398},
archivePrefix = {arXiv},
       eprint = {2403.05398},
 primaryClass = {astro-ph.IM},
       adsurl = {https://ui.adsabs.harvard.edu/abs/2024arXiv240305398M}
}

@ARTICLE{Roman_2022,
       author = {{Wang}, Yun and {Zhai}, Zhongxu and {Alavi}, Anahita and {Massara}, Elena and {Pisani}, Alice and {Benson}, Andrew and {Hirata}, Christopher M. and {Samushia}, Lado and {Weinberg}, David H. and {Colbert}, James and {Dor{\'e}}, Olivier and {Eifler}, Tim and {Heinrich}, Chen and {Ho}, Shirley and {Krause}, Elisabeth and {Padmanabhan}, Nikhil and {Spergel}, David and {Teplitz}, Harry I.},
        title = "{The High Latitude Spectroscopic Survey on the Nancy Grace Roman Space Telescope}",
      journal = {\apj},
         year = 2022,
        month = mar,
       volume = {928},
       number = {1},
          eid = {1},
        pages = {1},
          doi = {10.3847/1538-4357/ac4973},
archivePrefix = {arXiv},
       eprint = {2110.01829},
 primaryClass = {astro-ph.CO},
       adsurl = {https://ui.adsabs.harvard.edu/abs/2022ApJ...928....1W}
}

@INPROCEEDINGS{Lipman2023,
       author = {{Lipman}, Yaron and {Chen}, Ricky T.~Q. and {Ben-Hamu}, Heli and {Nickel}, Maximilian and {Le}, Matt},
        title = "{Flow Matching for Generative Modeling}",
      journal = {ICLR},
         year = 2023,
      booktitle = {Proceedings of ICLR},
          doi = {10.48550/arXiv.2210.02747},
archivePrefix = {arXiv},
       eprint = {2210.02747}
}

@INPROCEEDINGS{Tong2023,
       author = {{Tong}, Alexander and {Fatras}, Kilian and {Malkin}, Nikolay and {Huguet}, Guillaume and {Zhang}, Yanlei and {Rector-Brooks}, Jarrid and {Wolf}, Guy and {Bengio}, Yoshua},
        title = "{Improving and generalizing flow-based generative models with minibatch optimal transport}",
      journal = {TMLR},
      booktitle = {Proceedings of TMLR},
         year = 2024,
          url = {https://openreview.net/forum?id=CD9Snc73AW},
        issn = {2835-8856}
}

\begin{appendix} 

\section{Dimensionality of the power spectrum} \label{app: dimension}
Throughout this work, we employ spherically averaged (1D) power spectra for parameter inference. While the cylindrically averaged (2D) power spectrum retains directional information along the line-of-sight and perpendicular directions, which in principle encodes additional sensitivity to redshift-space distortions and ionization anisotropies, we find that the enhanced S/N achieved through spherical averaging yields superior inference performance in practice.

Table~\ref{tab:1d_vs_2d_comparison} presents a quantitative comparison of inference quality using 1D versus 2D power spectra for the fiducial survey configuration ($\mathrm{FOV}=100\mathrm{deg}^2$, $\sigma_z=0.001$, $M_\mathrm{h,min}=10^{11}\mathrm{M}_\odot$). We evaluate performance using the posterior volume (PV) metric, defined as the standard deviation of posterior samples normalized by the prior standard deviation, computed over 1000 test observations. Lower values indicate tighter constraints.

\begin{table}[h]
\centering
\caption{Comparison of 1D and 2D power spectrum performance for different data modes.}
\label{tab:1d_vs_2d_comparison}
\begin{tabular}{llcccc}
\toprule
\textbf{Dim.} & \textbf{Mode} & \textbf{PV} ($x_{\mathrm{HI}}$) & \textbf{PV}($\langle 1+\delta_\mathrm{HI}\rangle$) \\
\midrule
1D & 21cm & $0.0248 \pm 0.0015$ & $0.0677 \pm 0.0043$ \\
1D & Cross & $0.0893 \pm 0.0205$ & $0.1223 \pm 0.0234$ \\
1D & Both & $\fbox{$0.0176 \pm 0.0016$}$ & $\fbox{$0.0496 \pm 0.0032$}$ \\
\midrule
2D & 21cm & $0.0231 \pm 0.0080$ & $0.0769 \pm 0.0411$ \\
2D & Cross & $0.1758 \pm 0.0206$ & $0.2643 \pm 0.0203$ \\
2D & Both & $0.0193 \pm 0.0016$ & $0.0567 \pm 0.0054$ \\
\bottomrule
\end{tabular}
\tablefoot{Metrics are computed over the 1000 test samples with epistemic network uncertainties. Best values (lowest PV) are highlighted with boxes.}
\end{table}
For the 21cm auto-power spectrum, 1D and 2D performance are comparable, with 2D achieving modestly better constraints for $x_\mathrm{HI}$ ($\mathrm{PV}=0.023\pm0.008$ versus $0.025\pm0.002$ for 1D), though the difference is within the epistemic noise. However, for the 21cm galaxy cross-power spectrum, the pattern reverses dramatically. The 2D cross-power yields substantially degraded constraints ($\mathrm{PV}=0.176\pm0.021$ for $x_\mathrm{HI}$ versus $0.089\pm0.021$ for 1D, and $0.264\pm0.020$ versus $0.122\pm0.023$ for $\langle 1+\delta_\mathrm{HI}\rangle$). This behavior arises because the cross-correlation measurement is subject to additional uncertainty sources beyond those affecting the auto-power, and the directional information retained in the 2D format does not compensate for the loss of S/N from reduced mode averaging.

When combining both auto-power and cross-power measurements, the 1D version delivers the tightest constraints ($\mathrm{PV}=0.018\pm0.002$ for $x_\mathrm{HI}$ and $0.050\pm0.003$ for $\langle 1+\delta_\mathrm{HI}\rangle$), marginally outperforming the 2D combination ($\mathrm{PV}=0.019\pm0.002$ and 
$0.057\pm0.005$), though the difference is within the epistemic noise for $x_\mathrm{HI}$. The clear advantage of 1D is most pronounced for the cross-power alone, where the S/N gain from spherical averaging outweighs the loss of directional information. These results justify our choice to use 1D power spectra throughout the main analysis. 

\section{Epistemic uncertainty}\label{app: stochasticity}
In Section~\ref{sec: galaxy survey} we systematically varied the galaxy survey parameters and studied the constraining power of the 21cm galaxy cross-power spectrum by means of the posterior volume (PV). The PV metric carries an intrinsic uncertainty arising from the stochasticity of neural network training, which we quantify here. For each of the 27 survey configurations, we trained 6 networks with different random initializations, keeping all other hyperparameters fixed, amounting to 162 trained models in total. The same procedure was applied to the three models in Section~\ref{sec: EoR timeline}, where epistemic noise is small given the fiducial survey configuration, and to the models in Section~\ref{sec: astro}, where the combined cross+auto model shows the largest epistemic noise ($\mathrm{PV}=0.24\pm0.12$), likely due to the architectural challenges of training on two observables with vastly different information content. The resulting mean PV and standard deviation across runs are shown directly in Fig.~\ref{fig: posterior volume} and Fig.~\ref{fig:surveys}. We highlight that this comprehensive study is only feasible thanks to the computational efficiency of our SBI framework; traditional inference approaches would render such a large-scale exploration prohibitive. Furthermore, we verified that sampling uncertainties arising from the finite test set size are negligible compared to the epistemic noise reported here, confirming that network training stochasticity is the dominant source of uncertainty in the PV metric.
The variance in PV follows a clear hierarchy determined by the constraining power of the survey. Well-constrained configurations (e.g. $\mathrm{FOV}=100\mathrm{deg}^2$, $\sigma_z=0.001$, $M_\mathrm{h,min}=10^{10}\mathrm{M}_\odot$) yield $\mathrm{PV}=0.04\pm0.01$: the data are highly informative, the network learns consistently regardless of initialization, and the epistemic noise is low. At the other extreme, configurations with very high observational noise (e.g. $\sigma_z=0.1$) yield $\mathrm{PV}\approx1.0$
with negligible variance: the data are entirely uninformative and all initializations converge to the same uninformative posterior. The largest epistemic noise arises in intermediate configurations (e.g. $\mathrm{FOV}=100\mathrm{deg}^2$, $\sigma_z=0.01$, $M_\mathrm{h,min}=10^{11}\mathrm{M}_\odot$, $\mathrm{PV}=0.60\pm0.18$): here the data contain some information but the learning problem is hard enough that the network only succeeds from favorable initializations, leading to large run-to-run variance without further architecture tuning.

We attribute this behavior to two effects. First, networks trained on uninformative data face a harder optimization problem with loss curves more sensitive to initialization. Second, the network architecture and training hyperparameters were optimized for the fiducial survey configuration; applying them unchanged across all 27 configurations is an intentional design choice to isolate the effect of survey parameters, but it means the training is suboptimal for the most challenging configurations.
Overall, Fig.~\ref{fig:surveys} provides a clear qualitative hierarchy of survey performance that is robust across all training runs. However, particularly for intermediate configurations ($\sigma_z = 0.01$) where epistemic noise is largest, the individual PV values should not be interpreted as precise quantitative measurements, and conclusions drawn from these configurations should remain qualitative in nature.

\section{Saliency analysis}\label{app: saliency}
To understand which regions of the input power spectra most strongly influence
parameter inference by the neural network, we perform a saliency analysis by computing gradients of the log-probability with respect to the input features. This approach reveals where the network focuses its attention across different scales and redshifts, providing insight into which observational modes carry the most relevant information for constraining reionization parameters.

For a trained normalizing flow, the gradient $\nabla_{\mathbf{x}} \log p(\theta|\mathbf{x})$ quantifies the sensitivity of the posterior probability to small changes in each input feature. We compute the average absolute gradient across all test samples, creating saliency maps that highlight which $(k, z)$ bins in the power spectrum the network weights most heavily during inference. 

\begin{figure}[h!]
    \centering
    \includegraphics[width=0.49\textwidth]{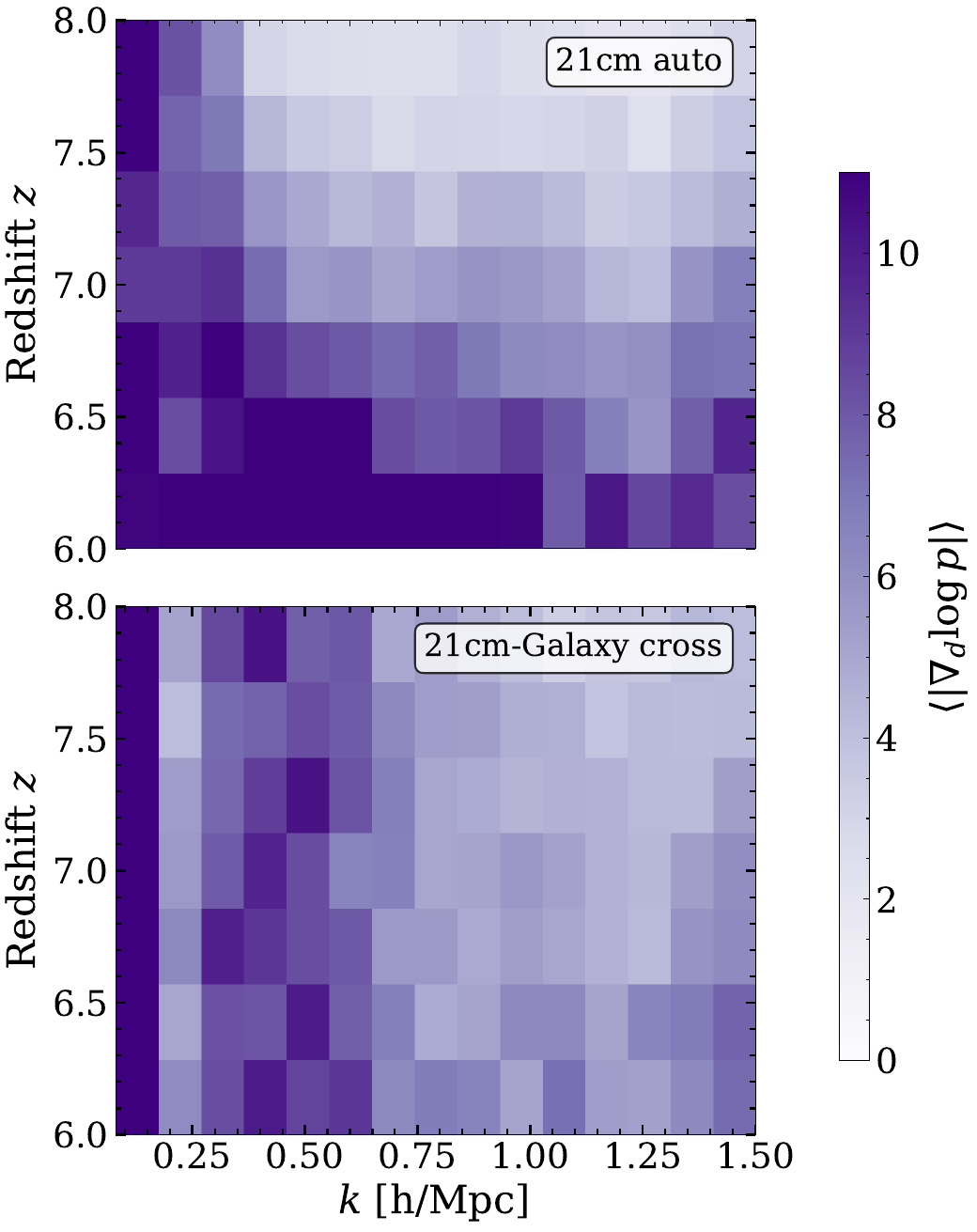}
    \caption{Saliency analysis for the inference of $x_\mathrm{HI}(z)$ and $\langle 1+\delta_\mathrm{HI}\rangle(z)$ from the 21cm auto-power spectrum (left) and the 21cm galaxy cross-power spectrum (right). The colormap shows the average absolute gradient of the log-probability with respect to the input power spectrum, indicating which redshift bins and spatial scales the network attends to most strongly. Both observables show high attention to the lowest $k$ bin across all redshifts, corresponding to the foreground-masked region that the network must learn to disregard. The 21cm auto-power focuses primarily on low redshifts across all scales, while the cross-power shows more uniform attention across redshifts but concentrates on intermediate scales ($k \sim 0.3$-$0.7\,h\mathrm{Mpc}^{-1}$), consistent with the characteristic sizes of ionized regions.}
    \label{fig:saliency}
\end{figure}

Fig.~\ref{fig:saliency} presents saliency maps for inference on $x_\mathrm{HI}(z)$ and $\langle 1+\delta_\mathrm{HI}\rangle(z)$ using the 21cm auto-power spectrum (upper panel) and the 21cm galaxy cross-power spectrum (lower panel). 

Both maps exhibit high gradient values at the lowest $k$ modes across all redshifts, corresponding to the foreground wedge region excluded by our observational mask. The network must learn to recognize and disregard this masked region, which contains no physical information about reionization, explaining the elevated attention in this regime.
Beyond the masked region, the two observables display distinct attention patterns. For the 21cm auto-power spectrum, the network concentrates attention on the lowest redshifts ($z \sim 6$-$7$) across a broad range of scales. This pattern reflects the fact that late-stage reionization, when the neutral fraction is rapidly declining and large-scale power is maximized, provides the strongest constraints on global reionization properties. In contrast, the 21cm galaxy cross-power spectrum shows relatively uniform attention across all redshifts but focuses on intermediate scales ($k \sim 0.3$-$0.7\,h\mathrm{Mpc}^{-1}$). These scales correspond to the characteristic sizes of ionized regions and the transition between regimes where galaxies trace ionized bubbles versus underlying density fluctuations, consistent with the physical expectations discussed in Section~\ref{sec: mutual info}.

These findings align with both the mutual information analysis presented in the main text and physical intuition about reionization signatures. The sensitivity of the cross-power to intermediate scales during all phases of reionization reflects its direct probe of ionization morphology, while the emphasis of the auto-power on late times captures the evolution of global neutral fraction. Nevertheless, we emphasize that these patterns are emergent properties of the trained network architecture and do not constitute a fundamental decomposition of information content. Alternative network designs or summary statistics might access the same underlying information through different pathways.

\section{Optimistic 21cm foregrounds}\label{app: optimistic}

Section~\ref{sec: astro} demonstrated that 21cm galaxy cross-power spectra provide tight constraints on astrophysical source properties when assuming deep galaxy surveys capable of detecting halos down to $M_\mathrm{h,min} = 10^{10}$ M$_\odot$ under moderate 21cm foreground avoidance. However, this requires detecting extremely faint galaxies that may lie beyond the reach of currently planned survey facilities. Here we show that optimistic 21cm foreground removal strategies offer an alternative pathway to achieving comparable constraints on source properties even with the more conservative halo mass threshold of $M_\mathrm{h,min} = 10^{11}$ M$_\odot$.

Under moderate foreground avoidance, we mask contaminated modes in the foreground wedge following the prescription of~\citet{21cmsense14}, which excludes modes below the horizon limit plus a conservative 0.1 $h$Mpc$^{-1}$ buffer. This approach sacrifices the large-scale modes most contaminated by foreground emission to ensure robust detection of the cosmological signal. However, recent advances in foreground removal techniques~\citep[e.g.,][]{Mertens_2020,acharya2024} suggest that more aggressive cleaning strategies may successfully recover these wedge modes, particularly for cross-correlation measurements where foreground contamination is partially suppressed through decorrelation between 21cm and galaxy fields. To explore this optimistic scenario, we retrain our inference framework assuming all modes down to the primary beam scale can be recovered (the "optimistic foreground model" of \texttt{21cmSense}). We apply this to the fiducial survey configuration (FOV $= 100$ deg$^2$, $\sigma_z = 0.001$, $M_\mathrm{h,min} = 10^{11}$ M$_\odot$) and perform inference on the four astrophysical source parameters $\{f_{\mathrm{esc},10}, f_{*,10}, \alpha_\mathrm{esc}, \alpha_*\}$.

\begin{figure*}[h]
\sidecaption
    \centering
    \includegraphics[width=12cm]{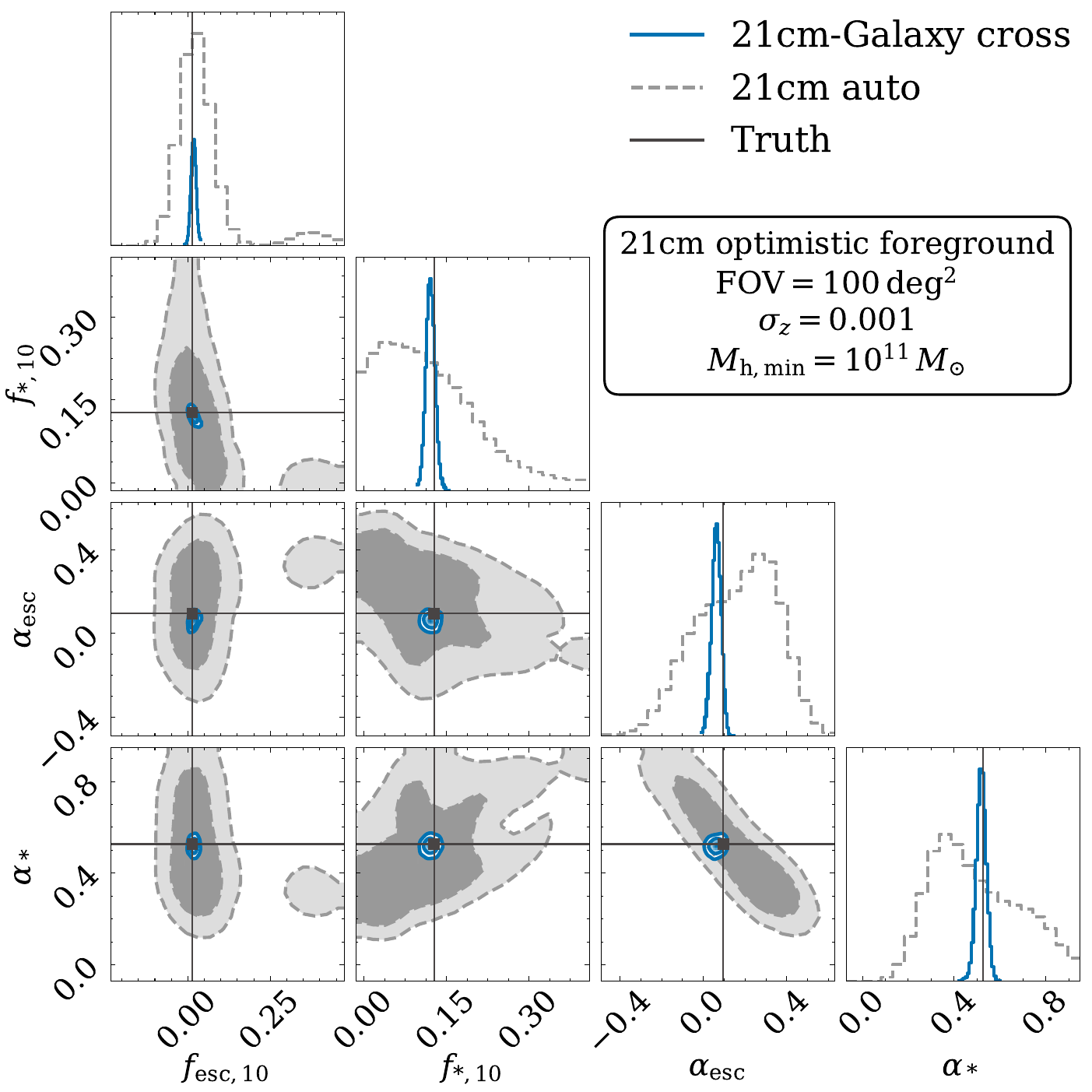}
    \caption{Inference of astrophysical source parameters from 21cm auto- and 21cm galaxy cross-power spectra for a randomly chosen sample assuming optimistic 21cm foreground removal. Corner plots show the marginalized posteriors for the four parameters controlling escape fraction and star formation efficiency: $\log_{10} f_{\mathrm{esc},10}$, $\log_{10} f_{*,10}$, $\alpha_\mathrm{esc}$, and $\alpha_*$. Blue, solid contours show cross-power constraints (R$^2 > 0.81$, posterior volumes $\sim 18$\% across the entire test dataset). Grey, dashed contours show auto-power constraints, which are virtually uninformative (R$^2$ < 0.51, posterior volumes $\sim 73$\% across the entire test dataset). Results assume the fiducial survey configuration with FOV $= 100$ deg$^2$, spectroscopic redshift uncertainty $\sigma_z = 0.001$, and halo mass detection threshold $M_\mathrm{h,min} = 10^{11}$ M$_\odot$, but with an optimistic 21cm foreground model that recovers wedge modes down to the primary beam scale. The dramatic improvement in cross-power constraints compared to moderate foregrounds with this mass threshold (Section~\ref{sec: astro}) demonstrates that recovering wedge modes can compensate for detecting fewer faint galaxies, offering a complementary observational pathway to constraining source properties. The persistent weakness of auto-power constraints reveals that additional 21cm modes do not break the fundamental degeneracies among source models that produce similar global ionization histories, highlighting the unique capability of cross-correlations to access source property information.}
\label{fig:astro_corner_opt}
\end{figure*}

Fig.~\ref{fig:astro_corner_opt} presents the inference results, revealing a dramatic improvement in source property constraints compared to the moderate foreground case with the same galaxy survey parameters. Across the test dataset, the 21cm galaxy cross-power spectrum (blue contours) achieves R$^2$ scores ranging from 0.81 to 0.86 across all four parameters, with $\mathrm{PV}=0.18\pm 0.04$ relative to the prior. These constraints are broadly comparable to, though somewhat degraded relative to, those obtained with $M_\mathrm{h,min} = 10^{10}$ M$_\odot$ under moderate foregrounds ($\mathrm{PV}=0.11\pm0.02$, Section~\ref{sec: astro}), demonstrating that recovering wedge modes can largely compensate for detecting fewer galaxies.
Critically, the 21cm auto-power spectrum (grey contours) shows no improvement under optimistic foregrounds, with R$^2$ scores remaining between 0.09 and 0.51 and $\mathrm{PV}=0.73\pm0.03$. This asymmetric improvement reveals a fundamental physical distinction. The auto-power spectrum is inherently insensitive to source properties because different combinations of escape fractions and star formation efficiencies can produce similar global ionization histories and thus nearly identical power spectrum shapes. Recovering additional modes does not break these degeneracies. In contrast, the cross-power spectrum directly probes the correlation between ionized regions and galaxy positions, which depends sensitively on how efficiently individual galaxies produce ionizing photons. The large-scale modes recovered by optimistic foreground removal are precisely those that encode the strongest source-bubble correlations, explaining why cross-power constraints improve substantially while auto-power constraints remain weak.

This result establishes two complementary pathways to constraining reionization source properties through 21cm galaxy cross-correlations. Deep galaxy surveys detecting faint sources ($M_\mathrm{h,min} \sim 10^{10}$ M$_\odot$) can achieve tight constraints even under conservative 21cm foreground treatment. Alternatively, surveys targeting brighter galaxies ($M_\mathrm{h,min} \sim 10^{11}$ M$_\odot$), which are more readily accessible to planned facilities such as the WST and PFS, can reach similar precision if paired with aggressive 21cm foreground cleaning that successfully recovers wedge modes. This flexibility is crucial for practical observational programs, as it allows constraints on source properties to be achieved through improvements in either the galaxy survey sensitivity or the 21cm foreground removal capability.
The dramatic contrast between cross-power and auto-power performance under optimistic foregrounds (R$^2 \sim 0.83$ versus R$^2 \sim 0.31$) underscores the unique role of cross-correlations in accessing source properties. While both observables benefit from increased sensitivity to the 21cm signal morphology, only the cross-power translates this into meaningful constraints on the parameters controlling where and how efficiently ionizing photons escape from galaxies. This reinforces the conclusion of Section~\ref{sec: astro} that 21cm galaxy cross-correlations provide not merely complementary information to auto-power measurements, but essential information for parameters that remain degenerate in single-probe analyses.

\end{appendix}

\end{document}